\documentclass[prd,twocolumn,superscriptaddress,nofootinbib,showpacs,epsfig,preprintnumbers,sort&compress]{revtex4-1}
\usepackage{hyperref}
\usepackage{graphicx}

\usepackage{color}
\usepackage{xcolor}
\usepackage{float}
\usepackage{graphicx}
\graphicspath{{../Figures_v7/}}
\usepackage{epstopdf}
\usepackage{graphics}
\usepackage[intlimits]{amsmath}
\usepackage[utf8]{inputenc}
\usepackage{esint}
\usepackage{svg}
\usepackage{dcolumn}
  \newcolumntype{d}[1]{D{.}{.}{#1}}

\newcommand{\hzdr}{\affiliation{Helmholtz-Zentrum Dresden-Rossendorf (HZDR), 01328 Dresden, Germany}}
\newcommand{\tudd}{\affiliation{Technische Universität Dresden, 01062 Dresden, Germany}}
\newcommand{\ific}{\affiliation{Instituto de Física Corpuscular, CSIC – Universidad de Valencia, 46980 Paterna, Spain}}
\newcommand{\vkta}{\affiliation{VKTA – Strahlenschutz, Analytik \& Entsorgung Rossendorf, 01328 Dresden, Germany}}
\newcommand{\gsi}{\affiliation{GSI Helmholtzzentrum für Schwerionenforschung, 06159 Darmstadt, Germany}}
\newcommand{\complu}{\affiliation{Grupo de Física Nuclear \& IPARCOS, Universidad Complutense de Madrid, 28040 Madrid, Spain}}
\newcommand{\ill}{\affiliation{Institut Laue-Langevin (ILL), 38042 Grenoble, France}}
\begin{document}

\title{Neutron flux and spectrum in the Dresden Felsenkeller underground facility studied by moderated $^3$He counters}

\author{M.~Grieger} \hzdr \tudd
\author{T.~Hensel} \hzdr \tudd
\author{J.~Agramunt} \ific
\author{D.~Bemmerer}\email[e-mail address:\,]{d.bemmerer@hzdr.de} \hzdr 
\author{D.~Degering} \vkta
\author{I.~Dillmann} \gsi
\author{L.M.~Fraile} \complu
\author{D.~Jordan} \ific
\author{U.~Köster} \ill
\author{M.~Marta} \gsi
\author{S.E.~Müller} \hzdr
\author{T.~Szücs} \hzdr
\author{J.L.~Taín} \ific
\author{K.~Zuber} \tudd

\date{\today}

%%%%%%%%%%%%%%%%%%%%%%%%%%%%%%%%%%%%%%%%%%%%%%%%%%%%%%%%%%%%%%%%%%%%%
\begin{abstract}

Ambient neutrons may cause significant background for underground experiments. Therefore, it is necessary to investigate their flux and energy spectrum in order to devise a proper shielding. Here, two sets of altogether ten moderated $^3$He neutron counters are used for a detailed study of the ambient neutron background in tunnel IV of the Felsenkeller facility, underground below 45\,m of rock in Dresden/Germany. One of the moderators is lined with lead and thus sensitive to neutrons of energies higher than 10\,MeV. For each $^3$He counter moderator assembly, the energy-dependent neutron sensitivity was calculated with the FLUKA code. The count rates of the ten detectors were then fitted with the MAXED and GRAVEL packages. As a result, both the neutron energy spectrum from 10$^{-9}$ to 300\,MeV and the flux integrated over the same energy range were determined experimentally.The data show that at a given depth, both the flux and the spectrum vary significantly depending on local conditions. Energy-integrated fluxes of $(0.61 \pm 0.05)$, $(1.96 \pm 0.15)$, and $(4.6 \pm 0.4) \times 10^{-4}$\,cm$^{-2}$\,s$^{-1}$, respectively, are measured for three sites within Felsenkeller tunnel IV which have similar muon flux but different shielding wall configurations. The integrated neutron flux data and the obtained spectra for the three sites are matched reasonably well by FLUKA Monte Carlo calculations that are based on the known muon flux and composition of the measurement room walls. 
\end{abstract}
%\begin{keyword}
%\end{keyword}
%%%%%%%%%%%%%%%%%%%%%%%%%%%%%%%%%%%%%%%%%%%%%%%%%%%%%%%%%%%%%%%%%%%%%

\maketitle

\section{Introduction}
\label{sec:Introduction}

Experiments striving for the lowest possible background in radiation detectors must be placed underground, in order to efficiently attenuate the direct and indirect effects of cosmic-ray-induced muons \cite{Heusser95-ARNPS,Formaggio04-ARNPS}. Underground experiments have proven particularly useful in the study of solar neutrinos \cite{Davis03-RMP,Gavrin18-SNC} and of neutrino flavor oscillations \cite{McDonald16-RMP,Kajita16-NobelLecture}.

In underground laboratories \cite{Bettini07-arxiv}, it is  important to precisely know the neutron background: In some cases, because neutrons are part of the solar neutrino-induced signal \cite{SNO02-PRL}, in other cases because they present a possible background for neutrinoless double-beta decay experiments \cite{Agostini18-PRL,Legend17-AIPCP,Zuber01-PLB} and dark matter searches \cite{Xenon18-PRL}, even though they cannot explain  \cite{Klinger15-PRL} a claimed dark matter detection \cite{Bernabei13-EPJC}. In underground nuclear astrophysics, the experimental study of the neutron source reactions for the astrophysical s-process \cite{Broggini18-PPNP} requires ultralow ambient neutron background. 

Despite the crucial importance of knowing, and then suppressing, the neutron background,  there is only a limited number of well-documented studies where both the flux and the energy spectrum of ambient neutrons were determined experimentally for underground laboratories \cite{Belli89-NCA,Arneodo99-NCA,Jordan13-APP,Jordan13-APP_Corr,Zhang14-PRD,Sonay18-PRC}. Other work, for example, concentrates on thermal neutrons \cite{Best16-NIMA,Debicki18-NIMA}, shows only one \cite{Niese07-JRNC} to three energy bins \cite{Rindi88-NIMA} or a limited energy range \cite{Du18-NIMA}, or does not present a deconvoluted neutron energy spectrum \cite{DaSilva95-NIMA}. It was reported that the flux of muon-induced neutrons seems to be underpredicted by GEANT4 Monte Carlo simulations with the standard physics list \cite{Du18-APP}, whereas FLUKA simulations seemed to fare better \cite{Kneissl19-APP}.

In addition, there are major discrepancies between studies that simply measure the neutron background at one given energy and site. An example is the thermal neutron flux in the Gran Sasso underground laboratory. Reported values vary by up to a factor of 6 between the four individual studies  \cite{Rindi88-NIMA,Belli89-NCA,Best16-NIMA,Debicki18-NIMA}, from $(0.32 \pm 0.09 \pm 0.04)\times10^{-6}$\,cm$^{-2}$\,s$^{-1}$ \cite{Best16-NIMA} up to $(2.05 \pm 0.06)\times10^{-6}$\,cm$^{-2}$\,s$^{-1}$ \cite{Rindi88-NIMA}. In a different deep-underground laboratory, Canfranc/Spain, the reported neutron flux \cite{Jordan13-APP} was recently revised upward by a factor of 4 \cite{Jordan13-APP_Corr}, further underlining the need for new and well-documented experimental efforts. 

The aim of the present study is to make a first step to address this unsatisfactory situation with a precisely documented measurement of the ambient neutron flux and energy spectrum in an underground laboratory, matched with a Monte Carlo simulation. The experiment was carried out at Felsenkeller in Dresden/Germany, which is shielded by 45\,m of hornblende monzonite rock \cite{Ludwig19-APP,Szucs19-EPJA}. 

In underground laboratories, there are two principal sources of ambient neutrons: first, neutrons that are produced in or near the experimental setup by cosmic-ray muons, here called ($\mu,n$) neutrons. Second, neutrons produced by ($\alpha,n$) reactions in the rock, with the $\alpha$ particles supplied by the $^{238}$U and $^{232}$Th decay chains\footnote{A weak contribution by the natural $^{235}$U decay chain is neglected here.} --- these are called ($\alpha,n$) neutrons here. Neutrons due to the spontaneous fission of $^{238}$U have similar characteristics as the ($\alpha,n$) neutrons but usually several orders of magnitude lower flux and are treated together with the ($\alpha,n$) neutrons here. Both ($\mu,n$) and ($\alpha,n$) neutrons show a maximum in the spectral flux near 1\,MeV neutron energy, as well as a similar overall spectral shape from thermal energies up to 10\,MeV. At even higher energies, $>$10\,MeV, the ($\alpha,n$) spectrum quickly drops, whereas the ($\mu,n$) spectrum extends to hundreds of MeV.

At a shallow depth such as the one of this study, it is expected \cite{Mei06-PRD, Kneissl19-APP} that ($\mu,n$) dominate over ($\alpha,n$) neutrons. It is noted that the opposite is true for deep-underground laboratories, where instead ($\alpha,n$) neutrons dominate. 

The present study was initially motivated by a project to install an ion accelerator and a second low background activity-counting facility \cite{Bemmerer18-SNC} in tunnels VIII and IX (Fig. \ref{fig:Map}) of Felsenkeller \cite{Ludwig19-APP,Szucs19-EPJA}. At the time of measurement, the tunnels hosting this new laboratory were not accessible, so the neighboring tunnel IV was studied instead, which presents very similar spatial characteristics. 

This work serves to complete the characterization of all background components in Felsenkeller. The measured and simulated muon flux \cite{Ludwig19-APP} and the measured count rate in large high-purity germanium detectors \cite{Szucs19-EPJA} were already reported elsewhere.

The present work is organized as follows.
The underground site is described in Sec. \ref{sec:Site}. FLUKA-based predictions for the neutron spectra and fluxes in three differently shielded rooms in this site are developed in Sec. \ref{sec:PredictedFlux}. 
A detector setup to address these expected fluxes is then developed, and its energy-dependent neutron sensitivity is calculated using FLUKA (Sec. \ref{sec:Sensitivities}). 
The experiment is described in \ref{sec:Experiment}.
Sec. \ref{sec:Fit} shows the obtained neutron fluxes and energy spectra. 
The data are compared with the literature in Sec. \ref{sec:Discussion}, and a summary and conclusions are offered in Sec. \ref{sec:Conclusion}.

%%%%%%%%%%%%%%%%%%%%%%%%%%%%%%%%%%%%%%%%%%%%%%%%%%%%%%%%%%%%%%%%%%%%%
\section{Underground site studied}
\label{sec:Site}

The shallow-underground site Felsenkeller is located in the Plauenscher Grund area along the Weißeritz river, in the southwestern corner of Dresden, Germany. Until the 18th century, there was a quarry, then one of the largest breweries of Germany was built there. The brewery closed in 1991, but nine storage tunnels constructed in 1856--1859 remain, as well as a number of overground buildings. The tunnels have horizontal access from the brewery courtyard and are interconnected in a comblike structure (Fig. \ref{fig:Map}). They are protected from cosmic rays by an overburden of 45\,m of hornblende monzonite rock \cite{Paelchen08-Book}. Rock samples taken inside the tunnels show specific activities of $130 \pm 30$ and $170 \pm 30$\,Bq/kg of $^{238}$U and $^{232}$Th, respectively \cite{Grieger16-Master}. Based on the measured vertical muon flux \cite{Ludwig19-APP}, the effective rock overburden is 140\,m water equivalent (m.w.e.).  Above the tunnels is a meadow planted with fruit trees.

The data for the present work were taken in tunnel~IV, at the following three different sites inside a \mbox{$\gamma$-counting} facility established in 1982 \cite{Helbig84-Isotopenpraxis}, and enlarged in 1995 \cite{Niese96-Apradiso}:
\begin{enumerate}
\item Messkammer 1 (hereafter called MK1) is shielded by 70\,cm serpentinite rock and 2\,cm pre-1945 steel  \cite{Helbig84-Isotopenpraxis} with an estimated total areal density of 160\,g/cm$^2$ \cite{Niese98-JRNC}. The serpentinite rock contains just 1.3 and 0.34\,Bq/kg of $^{238}$U and $^{232}$Th, respectively.
\item The shielding of Messkammer 2 (hereafter called MK2) consists of 1\,cm steel, 27\,cm iron pellets \mbox{($\rho = 4.5$\,g/cm$^3$)}, 3.5\,cm steel, 3\,cm lead, and 1.2\,cm steel, from the outside to the inside, in total 210\,g/cm$^2$ \cite{Niese96-Apradiso,Koehler09-Apradiso}. One part of the westernmost wall of MK2 was built with 1\,cm steel and 20\,cm lead instead, in total 235\,g/cm$^2$ for this part of the wall.
\item The workshop (hereafter called WS) is shielded from the surrounding hornblende monzonite rock by a 24\,cm thick brick wall.
\end{enumerate}

%%%%%%%%%%%%%%%%%%%%%%%%%%%%%%%%%%%%%%%%%%%%%%%%%%%%%%%%%%%%%%%%%%%%%
\begin{figure}
\includegraphics[width=\columnwidth,trim=1cm 0 0 0,clip]{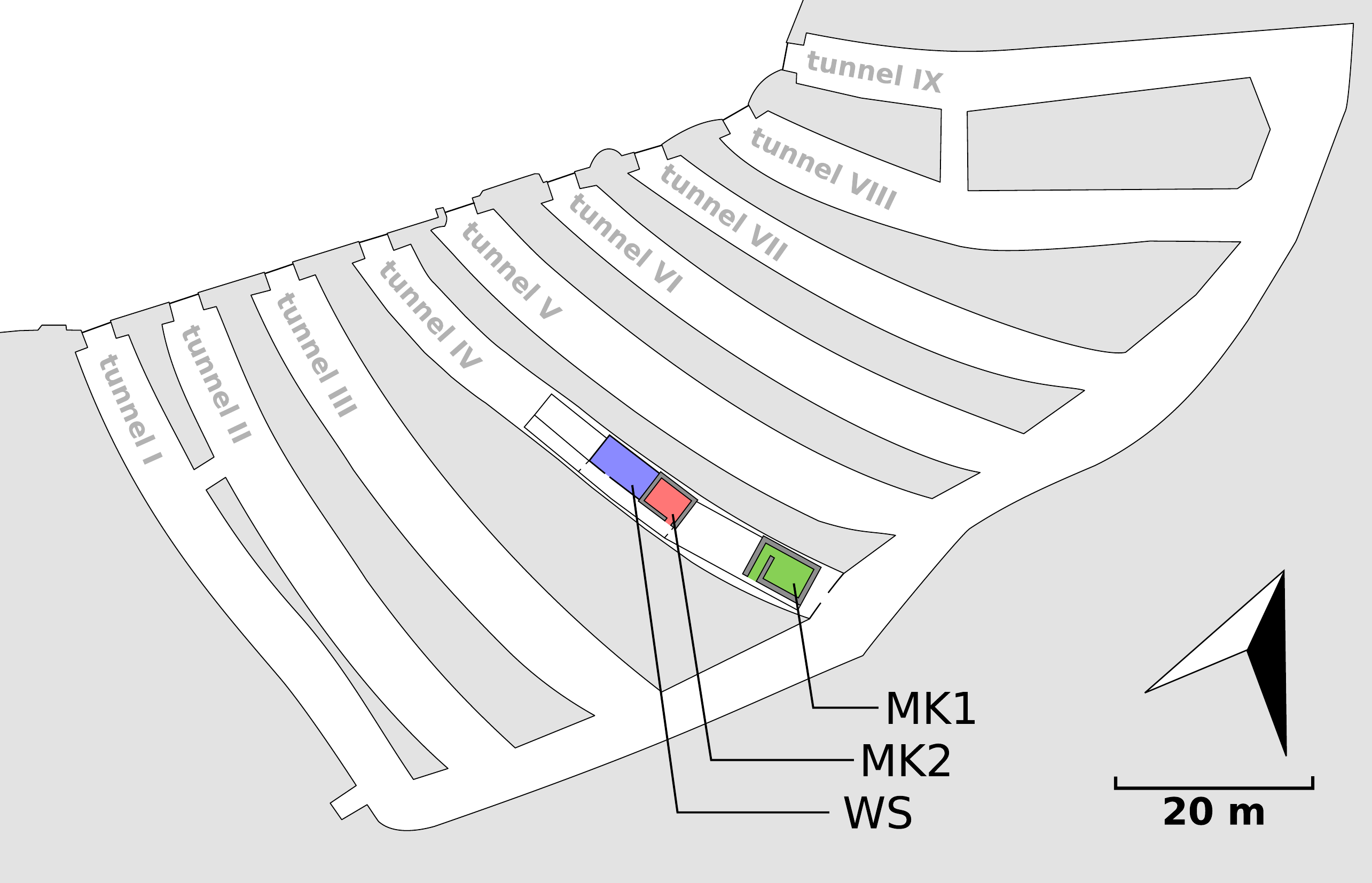}
\caption{\label{fig:Map} Map of the tunnels in the Felsenkeller underground site, Dresden/Germany. The arrow indicates geographic northern direction. For this work, the neutron flux was measured in tunnel IV at the three distinct locations WS, MK2, and inside of MK1. The external access to tunnels I--IX is from the west, where a courtyard and the overground buildings of the former brewery (not shown) are located.}
\end{figure}
%%%%%%%%%%%%%%%%%%%%%%%%%%%%%%%%%%%%%%%%%%%%%%%%%%%%%%%%%%%%%%%%%%%%%

All three sites are supplied by fresh, climatized air brought in by a ventilation duct from the outside. As a result, the radon concentration in the laboratory air is limited to 100--300\,Bq/m$^3$.

%%%%%%%%%%%%%%%%%%%%%%%%%%%%%%%%%%%%%%%%%%%%%%%%%%%%%%%%%%%%%%%%%%%%%
\section{Predicted neutron flux in Felsenkeller based on a FLUKA simulation}
\label{sec:PredictedFlux}

In order to guide the design of the experiment and the later fit of the experimental data, a FLUKA-based Monte Carlo prediction of the neutron flux and energy spectrum was developed. 

\subsection{General considerations}

Of the two components dominating the neutron flux in an underground laboratory, the flux of the first component, ($\mu,n$) neutrons, depends on the muon flux, hence the depth. The flux of the second component, ($\alpha,n$) neutrons (and spontaneous fission neutrons), does not depend on depth but on local conditions at the site studied.

The ($\mu,n$) neutrons may originate again from two different processes: either from the capture of stopped negative muons in the rock, or as part of a muon-induced hadronic shower. The relative importance of these two processes depends on muon energy and thus depth \cite{Mei06-PRD}. In a shallow laboratory such as the one studied here, stopped muon capture is expected to dominate, whereas the opposite is true deep underground \cite{Mei06-PRD}. 

The ($\alpha,n$) neutrons are created by $\alpha$ capture on light elements inside the rock, proportional to the specific activities of the natural $^{238}$U and $^{232}$Th decay chains \cite{Wulandari04-APP}. At great depth, where ($\mu,n$) neutrons are suppressed, ($\alpha,n$) neutrons dominate.

%%%%%%%%%%%%%%%%%%%%%%%%%%%%%%%%%%%%%%%%%%%%%%%%%%%%%%%%%%%%%%%%%%%%%
\begin{figure}[b]
\includegraphics[width=\columnwidth,trim=3mm 2mm 5mm 6.5mm]{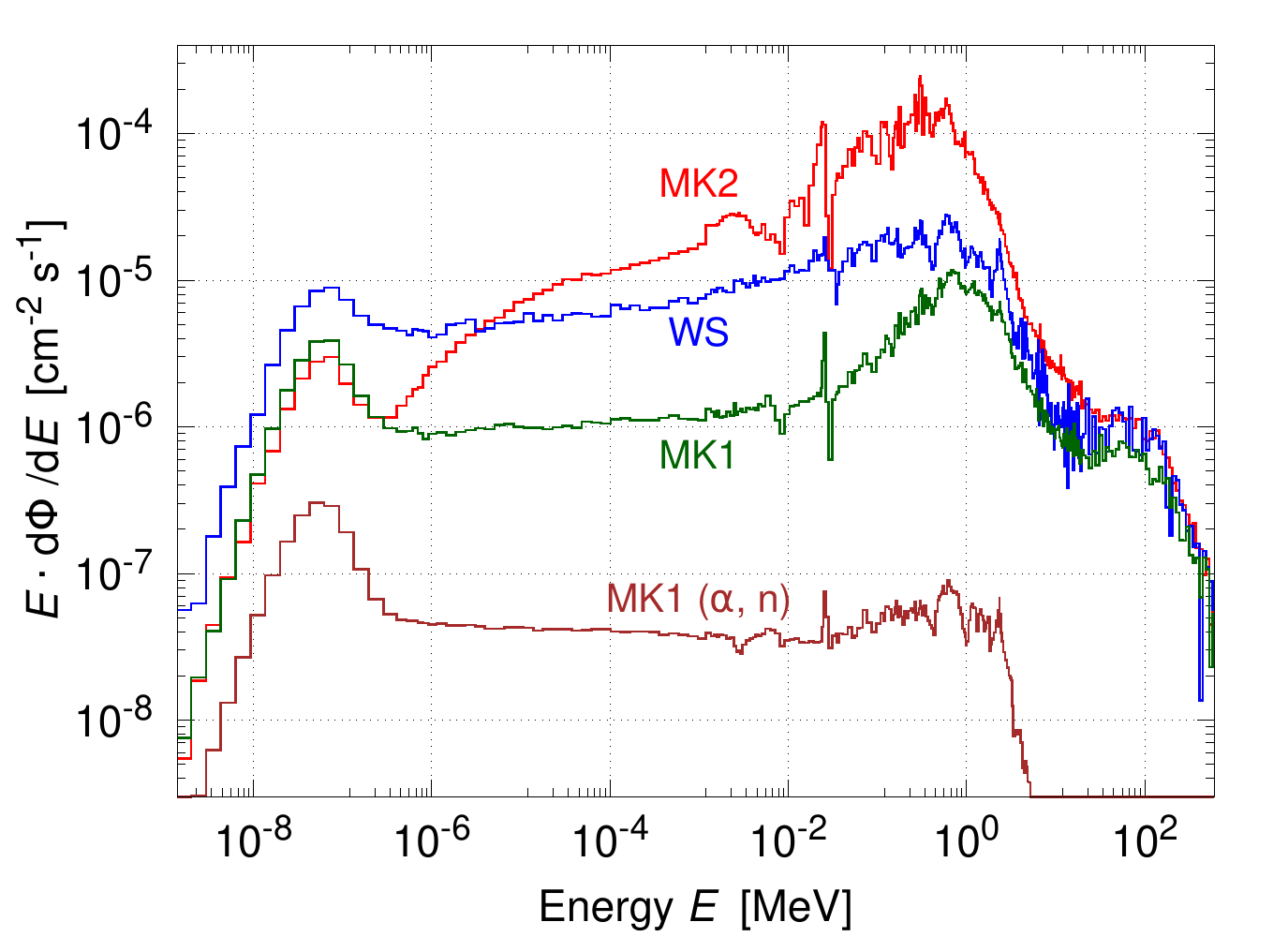}
\caption{\label{fig:SimulatedSpectra} Neutron energy spectra in the three Felsenkeller sites predicted by the FLUKA simulation: MK2 (red), WS (blue), and MK1 (green). All three are dominated by the ($\mu,n$) component. For reference, the weak ($\alpha,n$) contribution is shown separately for the case of MK1 (brown).}
\end{figure}
%%%%%%%%%%%%%%%%%%%%%%%%%%%%%%%%%%%%%%%%%%%%%%%%%%%%%%%%%%%%%%%%%%%%%

\subsection{Setup of the FLUKA simulation}

For the ($\mu,n$) neutrons, a simulation was performed using FLUKA \cite{Ferrari05-FLUKA,Boehlen14-NDS} version 2011.2x.6. An important physical input, namely the flux and angular distribution of cosmic-ray-induced muons at the sites studied, was recently measured \cite{Ludwig19-APP}, and these data are used here. Using the depth obtained from the measured \cite{Ludwig19-APP} muon flux, an average muon energy of 17\,GeV and a parametrized muon energy spectrum are adopted from the literature \cite{Mei06-PRD}. The rock composition is known from the geological literature \cite{Paelchen08-Book}, which was confirmed by the measured density of 2.75\,g/cm$^3$ of a rock sample taken. 

The wall compositions of the three sites MK1, MK2, and WS are known from construction records and confirmed by visual inspection. The water content of the rock is not known and may vary by season. For the present prediction, the hornblende monzonite rock and the MK1 serpentinite shielding were assumed to contain 3\% and 0.7\% (by mass) of water, respectively. 

In addition to the walls described above, there are also materials inside the rooms. For the sake of simplicity, only those materials inside the three rooms with nuclear charge $Z>20$ were included in the model. In particular, the high-purity germanium detectors in MK1 and MK2 are surrounded by lead castles. This lead may form a significant target for ($\mu,n$) neutron production and is included in the FLUKA model. 

A special consideration is necessary for the case of MK2. There, the iron and lead walls provide very little moderation. As a result, actually the detectors used here (see below, Sec. \ref{sec:Experiment}) affect the energy spectrum, but to first approximation not the energy-integrated flux. Therefore, for the case of the MK2 prediction, already the moderating effects of the detectors used here are included. 

In the FLUKA simulation, muons are started randomly on a virtual hemisphere that is placed inside the rock surrounding the tunnels, randomly distributed from 0.0 to 1.5\,m deep inside the rock. This value is sufficiently low that the experimental muon flux and angular distribution measured inside the tunnels \cite{Ludwig19-APP} can still be used. Test runs with 0.00--0.75\,m and 0--3\,m deep distributions instead did not significantly alter the neutron yields, showing that the presently assumed starting depth is robust.

For the ($\alpha,n$) neutrons, the known rock composition \cite{Paelchen08-Book} and the measured specific activities for the $^{232}$Th and $^{238}$U decay chains \cite{Grieger16-Master}, both assumed to be in secular equilibrium, are used. Then, the $\alpha$-induced neutron emission rate and spectrum are calculated with the NeuCBOT \cite{Westerdale17-NIMA} code and converted to an observed neutron flux using FLUKA. 

\subsection{Predicted neutron spectra}

For each of the three sites studied, the predicted spectrum is given by the sum of the ($\mu,n$) and ($\alpha,n$) contributions. The calculated spectra are given in the 296+100 FLUKA energy bins and smoothed to avoid unphysical effects in the later deconvolution (Fig. \ref{fig:SimulatedSpectra}). The ($\mu,n$) flux dominates at all energies. For comparison, the ($\alpha,n$) spectrum is shown separately for the case of MK1 (Fig. \ref{fig:SimulatedSpectra}, brown curve).
The energy integrals of the predicted fluxes are 5.8\,(4), 1.85\,(13), and 0.51\,(4) $\times$ 10$^{-4} \,\rm cm^{-2} s^{-1}$ for MK2, WS, and MK1, respectively. 

The FLUKA-simulated integrated neutron fluxes scale as 
$$ 11.2\,(12) : 3.6\,(4) : 1 \qquad \rm (MK2 : WS : MK1)$$
whereas the measured integrated muon fluxes \cite{Ludwig19-APP} in these three sites scale as 
$$1.28\,(14) : 1.49\,(15) : 1  \qquad \rm (MK2 : WS : MK1).$$

The strong differences found in the predicted energy-integrated neutron fluxes can therefore not be explained by the much lower differences in muon flux. The ($\alpha,n$) flux contributes at most 2\% to the total flux, again not enough to explain this effect.

Instead, it seems that the high-density shielding of MK2 presents a much more efficient target for ($\mu,n$) neutron production than the rock wall of MK1. 

In order to verify this hypothesis, a separate FLUKA simulation was carried out, again using 17\,GeV muons, and the literature \cite{Mei06-PRD} energy spectrum. Muons were incident from a spherical shell of 5\,m radius onto a sphere with 0.1\,m radius that consists of the material under study, which was given, in turn, by water, serpentinite, reinforced concrete, steel (74\% Fe, 18\% Cr, 8\% Ni), and lead. The neutrons were then detected at 1\,m distance from the smaller sphere. 

The results of this special FLUKA simulation are given in Table \ref{Table:mu_n}. The neutron yield in the major parts of the MK2 walls, steel and lead, is 7 and 15 times higher, respectively, than in serpentinite rock, the material of the MK1 walls, confirming the hypothesis.

At the highest neutron energies under study here, \mbox{$E = 10$--300\,MeV}, there is no contribution by ($\alpha,n$) neutrons, and the predicted flux is entirely given by ($\mu,n$) neutrons. Of the FLUKA-predicted neutrons in the  \mbox{10--300\,MeV} energy range, half originate in hadronic showers, and the other half from negative muon capture. At lower neutron energies, \mbox{$E=10^{-9}$--1\,MeV}, only 20\% of the predicted ($\mu,n$) neutrons are from showers, and 80\% from capture.

For ($\mu,n$) neutrons from hadronic showers, there may in principle be more than one neutron detected within the time window of the data acquisition system. The FLUKA simulation shows that these effects contribute less than 2\% to the neutron count rate. 

%%%%%%%%%%%%%%%%%%%%%%%%%%%%%%%%%%%%%%%%%%%%%%%%%%%%%%%%%%%%%%%%%%%%%
\begin{table}[bt]
\centering
\resizebox{\columnwidth}{!}{%
\begin{tabular}{|l|d{7}|d{7}|d{7}|d{7}|}
\hline
\multicolumn{1}{|l|}{~~Material}  & \multicolumn{4}{c|}{($\mu,n$) yield, relative to water}     \\
\multicolumn{1}{|l|}{}            & \multicolumn{1}{c|}{All energies}   & \multicolumn{1}{c|}{10$^{-9}$--10$^{-6}$} & \multicolumn{1}{c|}{10$^{-6}$--10} & \multicolumn{1}{c|}{10--300\,MeV}\\ \hline \hline
Water                & 1 &   0.102\,(1)   & 0.726\,(4)   & 0.171\,(2)    \\
Serpentinite         & 3.124\,(13) &   0.218\,(4)   & 2.520\,(12)  & 0.377\,(3)    \\
Reinf.\,concrete     & 3.511\,(15) &   0.011\,(1)   & 3.086\,(7)   & 0.401\,(3)    \\
Steel 				 & 21.2\,(2) &   0.0	       & 19.9\,(2)   & 1.25\,(2)   \\
Lead                 & 46.5\,(5) &   0.0          &  44.6\,(5) & 1.94\,(2)   \\ \hline
\end{tabular}
}%resizebox
\caption{Yield for muon-induced neutrons in selected materials from FLUKA. See text for details.}
\label{Table:mu_n}
\end{table}
%%%%%%%%%%%%%%%%%%%%%%%%%%%%%%%%%%%%%%%%%%%%%%%%%%%%%%%%%%%%%%%%%%%%%

%%%%%%%%%%%%%%%%%%%%%%%%%%%%%%%%%%%%%%%%%%%%%%%%%%%%%%%%%%%%%%%%%%%%%
%\section{Calculated detector sensitivities from FLUKA}
\section{Detection system design and calculated neutron sensitivity}
%\section{Monte Carlo guided detector design and evaluation of sensitivity}
\label{sec:Sensitivities}

This section is devoted to the calculation of the neutron sensitivities of the detectors used. First, some general concepts to study isotropic neutron fluxes are recalled for reference (Sec. \ref{subsec:General}). 
Then, a physical motivation for the choice of the detector setup is given (Sec. \ref{subsec:DetectorSetup}).  

The detector sensitivities are then calculated for isotropic neutron fluxes for all detectors (Sec. \ref{subsec:FLUKA}). Finally, possible angular effects are studied (Sec. \ref{subsec:Anisotropy}). 

\subsection{General considerations}
\label{subsec:General}

It is expected that the present neutron flux is dominated by negative muon capture \cite{Mei06-PRD} and therefore approximately isotropic (Sec. \ref{sec:PredictedFlux}). Therefore, some concepts developed previously for studying the isotropic neutron flux in nuclear reactors \cite{Fermi46-Pile} may be applied here. For the present purposes, the neutron flux density $\Phi$ is taken as 
\begin{equation}
\Phi = n_n v \label{eq:Fluxdef}
\end{equation}
where $n_n$ is the number density of neutrons, and $v$ their average velocity. $\Phi$ is referred to as the "neutron flux," measured in units cm$^{-2}$\,s$^{-1}$. Each detector is characterized by its neutron sensitivity $S$, given by
\begin{equation} \label{eq:Sensitivity}
R = S \Phi
\end{equation}
where $R$ is the count rate, in units s$^{-1}$. The sensitivity is measured in units cm$^2$. For the case of a tube counter, where the counter length $l$ is much larger than the diameter $d$, the sensitivity and side surface area $A = \pi ld$ of the tube are connected by \cite{Khokonov10-PAN}
\begin{equation} \label{eq:Four}
S = \frac{R}{\Phi} = \frac{A \varepsilon j_+}{\Phi} = A \varepsilon \frac{j_+}{\Phi} = \frac{A \varepsilon}{4}
\end{equation} 
where $\varepsilon$ is the efficiency for the counter to detect a neutron that passes its side-surface area, and 
\begin{eqnarray} \label{eq:Current}
j_+ & = & \frac{1}{4 \pi} \int\limits_0^{2\pi}d\phi \int\limits_0^{\infty}dr \int\limits_0^{\pi/2}d\theta \frac{\Phi}{\lambda_s} e^{-r/\lambda_s} \frac{|\cos \theta|}{r^2} r^2 \sin\theta  \nonumber \\
 & = & \frac{\Phi}{4}
\end{eqnarray}
is the directed current which passes through a unit surface area \cite{Soodak50-Book}, with $\lambda_s$ the mean free path for neutrons in the medium surrounding the detector. For a typical $^3$He counter with 1''  diameter and a few atmospheres gas pressure, $\varepsilon \sim$ 1 for thermal neutrons. 
In the case of a nonmonochromatric flux, Eq.\:(\ref{eq:Sensitivity}) becomes
\begin{equation} \label{eq:SpectralSensitivity}
R = \int\limits_{0}^{\infty} S(E) \Phi(E) dE 
\end{equation}
with $\Phi(E)$ the neutron flux per unit energy interval \cite{Lamarsh18-Book}.

When combining Eqs. (\ref{eq:Sensitivity}) and (\ref{eq:Four}), the neutron flux $\Phi$ is then given by 
\begin{equation} \label{eq:FluxAnalysis}
\Phi = \frac{R}{S} = \frac{4R}{\varepsilon A}.
\end{equation}

These relations are needed when sensitivity data are compared to experiments with a neutron beam \cite[e.g.]{Beyer12-JINST,Beyer18-EPJA}, e.g., where typically the side surface area times efficiency $A\varepsilon$ are measured. 

%%%%%%%%%%%%%%%%%%%%%%%%%%%%%%%%%%%%%%%%%%%%%%%%%%%%%%%%%%%%%%%%%%%%%
\subsection{Detector setup adopted}
\label{subsec:DetectorSetup}

In order to detect neutrons of higher energies than thermal and epithermal, in addition to a bare $^3$He counter, it is necessary to use an array of $^3$He detectors surrounded with polyethylene moderators. As a rule of thumb, it is expected that the energy of the highest neutron sensitivity shifts toward higher energy with increasing amount of moderator, while the overall sensitivity decreases somewhat. As a result, by using several different moderator sizes, the neutron energy spectrum may be determined.

Above 10\,MeV neutron energy, further increases of the moderator size are impractical, because the decrease in overall sensitivity becomes severe. Instead, a lead-lined polyethylene moderator is needed. The lead acts as a neutron multiplier by way of of ($n,xn$) reactions (with $x \in \{2,3,4,5,...\}$), once the neutron energy exceeds the respective reaction threshold of $E_{{\rm threshold},x}$ = 7.4, 14.2, 22.3, 29.1\,MeV, respectively, for $x$ = 2--5 secondary neutrons.

In order to enable a physically meaningful fit of the count rates by a neutron energy spectrum, the detector assemblies have been designed so that the structures in the predicted neutron flux (Fig.\:\ref{fig:SimulatedSpectra}, Sec.\:\ref{sec:PredictedFlux}) are matched with roughly similar structures in the energy-dependent neutron sensitivities. The adopted array of detector assemblies includes moderators ranging in size from 4.5 to 27.0\,cm (Table \ref{Table:Detectors}). In particular,

\begin{itemize}
\item The bare detector A0 addresses the thermal peak expected for all three locations (Fig. \ref{fig:SimulatedSpectra}). 
\item Assembly A3 matches the downscattered ($\mu,n$) peak expected in all sites at 0.3\,MeV. 
\item The difference of assemblies B8 and B9 addresses the ($\mu,n$) peak expected around 100\,MeV (Fig. \ref{fig:SimulatedSpectra}). 
\item Assemblies A2 and B7 with their very wide sensitivity pattern match the flat spectrum from 10$^{-6}$ to 10$^{-1}$\,MeV expected at all three sites.
\end{itemize}

%%%%%%%%%%%%%%%%%%%%%%%%%%%%%%%%%%%%%%%%%%%%%%%%%%%%%%%%%%%%%%%%%%%%%
\begin{table}
\begin{tabular}{|c|rrr|c|}
\hline
~~Detector~~ & \multicolumn{3}{c|}{~~Moderator size [cm]~~} & Remarks \\ 
 & ~~Height & ~~Width & \multicolumn{1}{c|}{~~Length} & \\ \hline \hline
A0 & \multicolumn{1}{c}{~~~~-} & \multicolumn{1}{c}{~~~~-} & \multicolumn{1}{c|}{~~-} & Unmoderated \\
A1 & 4.5~ & 4.5 & 70.0~ &  \\
A2 & 7.0~ & 7.0 & 70.0~ &  \\
A3 & 12.0~ & 12.0 & 70.0~ &  \\
A4 & 18.0~ & 18.0 & 70.0~ &  \\
A5 & 22.5~ & 22.5 & 70.0~ &  \\
A6 & 27.0~ & 27.0 & 70.0~ &  \\ \hline \hline
B7 & 7.0~ & 7.0 & 40.5~ &  \\
B8 & 22.5~ & 22.5 & 40.5~ &  \\
B9 & 21.0~ & 21.0 & 40.5~ & 0.5\,cm lead liner\\ \hline
\end{tabular}
\caption{\label{Table:Detectors} Dimensions of the polyethylene moderators used together with the $^3$He counters in campaigns A (A0--A6) and B (B7--B9), respectively. See text for details.}
\end{table}
%%%%%%%%%%%%%%%%%%%%%%%%%%%%%%%%%%%%%%%%%%%%%%%%%%%%%%%%%%%%%%%%%%%%%

%%%%%%%%%%%%%%%%%%%%%%%%%%%%%%%%%%%%%%%%%%%%%%%%%%%%%%%%%%%%%%%%%%%%%
\subsection{Evaluation of sensitivity calculation}
\label{subsec:FLUKA}

\begin{figure}
\includegraphics[width=\columnwidth,trim=4mm 0 6mm 1.5mm]{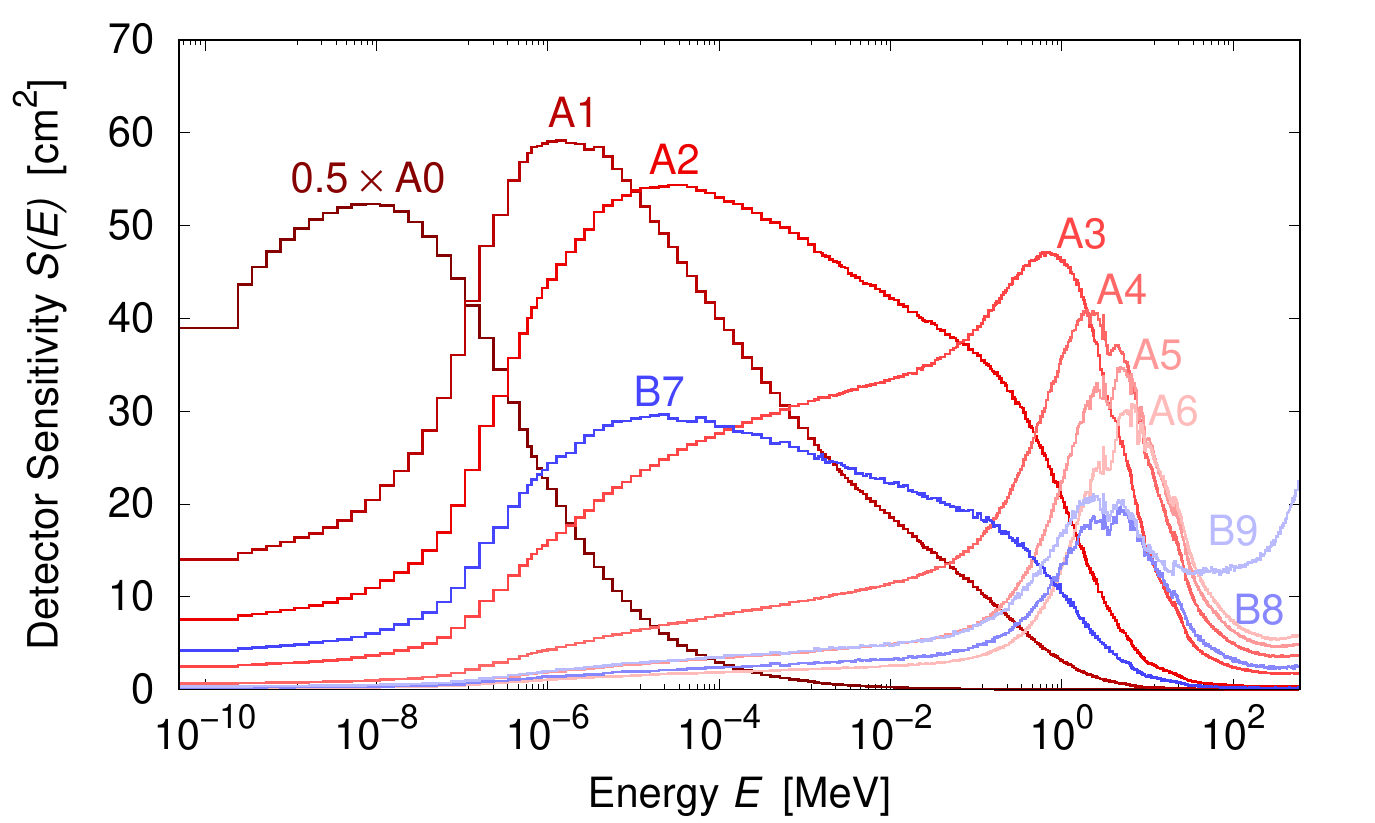}
\caption{\label{fig:Sensitivity} Energy-dependent neutron sensitivity $S(E)$, calculated by FLUKA for each detector-moderator assembly. See Table \ref{Table:Detectors} for details on the detector assemblies. For  A0 (bare $^3$He detector), $S(E)$ was scaled by a factor of 0.5 for the presentation.
}
\end{figure}

As a first step for the sensitivity calculation, the geometry of each $^3$He tube and its moderator were modeled with FLUKA \cite{Ferrari05-FLUKA,Boehlen14-NDS}, 2011.2x.6. Subsequently, the energy-dependent neutron sensitivity $S(E)$ was calculated for an isotropic neutron flux, in 396 energy bins covering the range from $5 \times 10^{-11}$ to 600\,MeV. Each assembly was simulated separately in vacuum, and neutrons were assumed to come in isotropically from the surface of a spherical shell of 50\,cm radius. The results of the simulation are shown in Fig. \ref{fig:Sensitivity}.

The maximum sensitivity for the bare $^3$He detector from the FLUKA simulation was $S(7.5\times10^{-9}\,{\rm MeV})$ = \linebreak 105\,cm$^2$. Since the detector was used with 10\,bar working pressure, this number cannot be directly compared with the data sheet value of $S$ = 144\,cm$^2$ for the 20\,bar case. Using analytical expressions given in Ref.\:\cite{Khokonov10-PAN}, \mbox{$S$ = 110\,cm$^2$} is found in the single-velocity approximation, matching the FLUKA result within 5\%. 

When comparing the calculated sensitivity curves $S(E)$ for assemblies A0--A6, it is observed that the energy of the peak sensitivity, which is near thermal energies for the bare detector A0, shifts step by step to higher energies. This behavior continues up to assemblies A5 and A6, with a peak sensitivity at $E\sim5$\,MeV. 

In order to also address energies $E>$ 10\,MeV, the lead-lined assembly B9 is used. Over a wide energy range 10$^{-6}$\,MeV $< E <$ 10\,MeV, its sensitivity curve is similar to assembly B8. For $E>$ 10\,MeV, the sensitivity of assembly B9 rises due to its 0.5\,cm lead liner. 

In order to study the mutual effect of neighboring detector assemblies on each other, for each case the sensitivity calculation was repeated adding the two neighboring detector assemblies as passive materials in the simulation. For assemblies A0 and A1, this causes the peak sensitivity to shift slightly to higher energies. For the other assemblies the sensitivity is reduced by up to 5\%.

%%%%%%%%%%%%%%%%%%%%%%%%%%%%%%%%%%%%%%%%%%%%%%%%%%%%%%%%%%%%%%%%%%%%%
\subsection{Study of angular distribution effects}
\label{subsec:Anisotropy}

The calculated sensitivities (Fig.\:\ref{fig:Sensitivity}, Sec.\:\ref{subsec:FLUKA}) were obtained for isotropic flux, but the $^3$He tubes included in the assemblies have a cylindric geometry, leading to a possible sensitivity to a nonisotropic angular distribution of the incident neutron flux. 

The largest such anisotropy is expected for the case of MK1. There, instead of the laboratory walls, the lead castles of the high-purity germanium (HPGe) detectors form the principal ($\mu,n$) target, meaning that there will be more neutrons hitting the assemblies from the sides than from above or below. 

Therefore, the sensitivity calculation was repeated for MK1 with the actual simulated neutron angular distribution, as obtained by FLUKA, fed into the FLUKA simulation of the sensitivity. It was found that for assemblies A4--A6, which are most sensitive to the relevant neutron energy range for unmoderated ($\mu,n$) neutrons, the sensitivity increased by 3.6--3.8\%. For all other detectors at MK1 and for all assemblies at the other two sites WS and MK2, the relative effect was below $\leq$2\%.

In order to take this effect into account, 4\% additional systematic uncertainty is added to the error budget.

%%%%%%%%%%%%%%%%%%%%%%%%%%%%%%%%%%%%%%%%%%%%%%%%%%%%%%%%%%%%%%%%%%%%%
\section{Experiment}
\label{sec:Experiment}

In the experiment, two sets of $^3$He-filled ionization chambers were used to determine both the total neutron flux and the energy spectrum. These counters are based on the $^3$He(n,p)$^3$H nuclear reaction, which has a $Q$ value of 764\,keV. The $^3$He(n,p)$^3$H cross section is ($5333 \pm 7$)\,barn for thermal neutrons and follows the 1/$v$ law (where $v$ is the neutron velocity) over a wide range of energies \cite{Mughabghab06-Book}. 

In order to check the performance of the detectors, runs with two different $^{252}$Cf spontaneous fission neutron sources with known activities were recorded prior to each measurement campaign. The $^{252}$Cf runs also serve as a reference for the shape of the neutron-induced pulse height spectrum.

%%%%%%%%%%%%%%%%%%%%%%%%%%%%%%%%%%%%%%%%%%%%%%%%%%%%%%%%%%%%%%%%%%%%%
\subsection{Experimental campaign A}
\label{subsec:CampaignA}

For campaign A, the detectors and moderators of the previous Canfranc neutron flux measurement  \cite{Jordan13-APP,Jordan13-APP_Corr} are used again here, but with 10\,bar instead of previously 20\,bar working pressure. An additional, bare $^3$He counter monitors the thermal neutron flux. 

The campaign A detectors are seven $^3$He proportional counter tubes of type LND-252248\footnote{LND Inc., Oceanside, New York, USA.}. The tubes have an active length and diameter  of 60.0 and 2.44\,cm, respectively, and are filled with 10\,bar working gas (97\% $^3$He, 3\% CO$_2$). The tubes were on loan from the BELEN / BRIKEN experiment \cite{Tarifeno17-JINST} which uses a large array of moderated $^3$He tubes to study $\beta$-delayed neutron emission \cite{Caballero18-PRC}. Here, the detector-moderator assemblies are called A0--A6, respectively (Table \ref{Table:Detectors}).

Each tube was connected to one input channel of one of two 16-fold Mesytec MPR16-HV preamplifiers and supplied with 1450\,V high voltage. For each preamplifier, only the first group of four channels was used, in order to minimize the length of the cables and limit the noise. The differential output of each preamplifier channel used was sent to a Mesytec STM16+ shaping amplifier and then to a Struck SIS3302 VME 16-bit, 100\,MS/s sampling digitizer. In order to determine the dead time of the system, a 10\,Hz pulse generator signal was fed into each preamplifier channel. The digitizer self-triggered separately for each channel and passed the timestamped signal to the GASIFIC data acquisition system \cite{Agramunt16-NIMA}, which then saved it to disk for offline analysis.

This experimental setup, including the data acquisition chain, was designed to be as similar as possible to the one used previously in Canfranc/Spain \cite{Jordan13-APP,Jordan13-APP_Corr}. The only differences were, first, the $^3$He gas pressure of 10\,bar here instead of 20\,bar in Canfranc, and, second, the unmoderated counter A0 added in the present experiment.

%%%%%%%%%%%%%%%%%%%%%%%%%%%%%%%%%%%%%%%%%%%%%%%%%%%%%%%%%%%%%%%%%%%%%
\begin{figure*}
\includegraphics[width=1.00\textwidth]{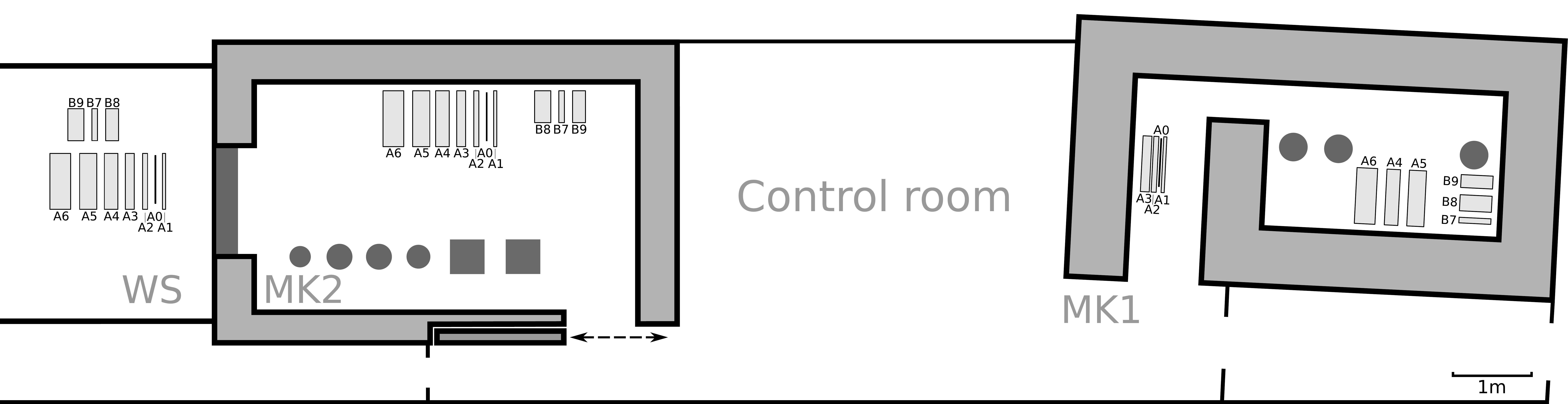}
\caption{%
\label{fig:Locations} Sketch of the three locations studied inside tunnel IV, with the approximate locations of the $^3$He detector-moderator assemblies shown. Location MK2 includes a heavy sliding door and has one side filled with two rows of lead blocks instead of its shielding wall. Location WS is closest to the tunnel entrance. The lead castles included in MK2 and MK1 are shown as circles. See text for details.
}
\end{figure*}
%%%%%%%%%%%%%%%%%%%%%%%%%%%%%%%%%%%%%%%%%%%%%%%%%%%%%%%%%%%%%%%%%%%%%

Each of the three sites MK2, WS, and MK1 \mbox{(Fig.\:\ref{fig:Locations})} was studied in turn. Counting times of 28.7, 12.5 and 8.6\,days were performed in the periods 15.12.2014--16.01.2015, 16.01.--03.02.2015, and 03.02.--12.02.2015, respectively. The measured dead time was always $\leq$0.3\%. 

At each site, the individual $^3$He tube-detector assemblies were placed parallel to each other, with typically 9\,cm of open space between them (Fig.\:\ref{fig:Locations}). This placement was due to electronic noise considerations, which demand short cable lengths. The star-shaped configuration previously adopted in Canfranc \cite{Jordan13-APP,Jordan13-APP_Corr} was not possible here due to space constraints. For the same reason, in the case of MK1, the four smallest assemblies A0--A4 had to be placed in the labyrinth area near the entrance of MK1, so that they were surrounded by the serpentinite only on five out of six sides.

The two measurement bunkers MK1 and MK2 contain three and six lead-shielded HPGe detectors, respectively, that are used for radioactivity measurements. MK2 has a heavy sliding door, and the present data acquisition was stopped typically 150\,minutes per day for maintenance and sample changes in the HPGe detectors. In MK1 and WS, the data acquisition was running continuously.

Prior to starting the ambient neutron measurement at each of the three sites, data were recorded with a $^{252}$Cf spontaneous fission neutron source placed centrally on top of each detector-moderator assembly, in turn. The neutron emission rate of the $^{252}$Cf source used for campaign A had been determined by Physikalisch-Technische Bundesanstalt\:(PTB), Braunschweig, Germany. Extrapolated to the time of the current measurement, it was $(6800 \pm 110)$\,s$^{-1}$. 

%%%%%%%%%%%%%%%%%%%%%%%%%%%%%%%%%%%%%%%%%%%%%%%%%%%%%%%%%%%%%%%%%%%%%
\subsection{Experimental campaign B}
\label{subsec:CampaignB}

Campaign B concentrated on the neutron flux above 10\,MeV neutron energy. Three $^3$He counter tubes of type LND-252189\footnote{Again LND Inc., Oceanside, New York, USA.}, on loan from Institut Laue-Langevin, Grenoble, France, were used. The tubes have an active length of 30.5\,cm, an outer diameter of 2.54\,cm, and are filled with 10\,bar working gas (97\% $^3$He, 3\% CO$_2$). The three detector-moderator assemblies are called B7--B9, respectively  (Table \ref{Table:Detectors}).

Assembly B7 was designed to resemble detector A2, in order to facilitate the connection between campaigns A and B. Detector assemblies B8 and B9 have similar polyethylene moderator sizes, but B9 is additionally fitted with a 0.5\,cm thick lead liner included at 5\,cm depth in the polyethylene matrix.

The tubes were operated at 1900\,V. For each one, the signal was amplified by a Mesytec MRS-2000 preamplifier and an Ortec 671 spectroscopy amplifier, then passed to a histogramming Ortec EtherNIM 919E analog-to-digital converter and multichannel buffer unit. The dead time, as obtained by the Gedcke-Hale algorithm \cite{Jenkins81-Book} implemented in the 919E unit, was $\leq$0.2\%. The pulse height spectrum was saved every 30\,minutes on hard disk for later analysis. 

As in campaign A, also in campaign B, the assemblies were placed subsequently in the three sites MK2, MK1, and WS (Figure \ref{fig:Locations}). Data were taken for periods of 10.6, 25.3, and 14.0\,days in the periods 26.08.--09.09.2016, 09.09.--13.10.2016, and 13.10.--27.10.2016, respectively. The first three days of the MK2 campaign were excluded from the analysis due to excessive electronics noise. In all three sites, data taking was never interrupted, even during the daily maintenance and sample change periods in MK2. 

For campaign B, the $^{252}$Cf benchmark measurements were performed at the HZDR Rossendorf surface site, using a $^{252}$Cf source with a neutron emission rate of \mbox{$(5200 \pm 800)$\,s$^{-1}$}, corrected for the time of the measurement.

%%%%%%%%%%%%%%%%%%%%%%%%%%%%%%%%%%%%%%%%%%%%%%%%%%%%%%%%%%%%%%%%%%%%%
\begin{figure*}
\includegraphics[width=0.5\textwidth]{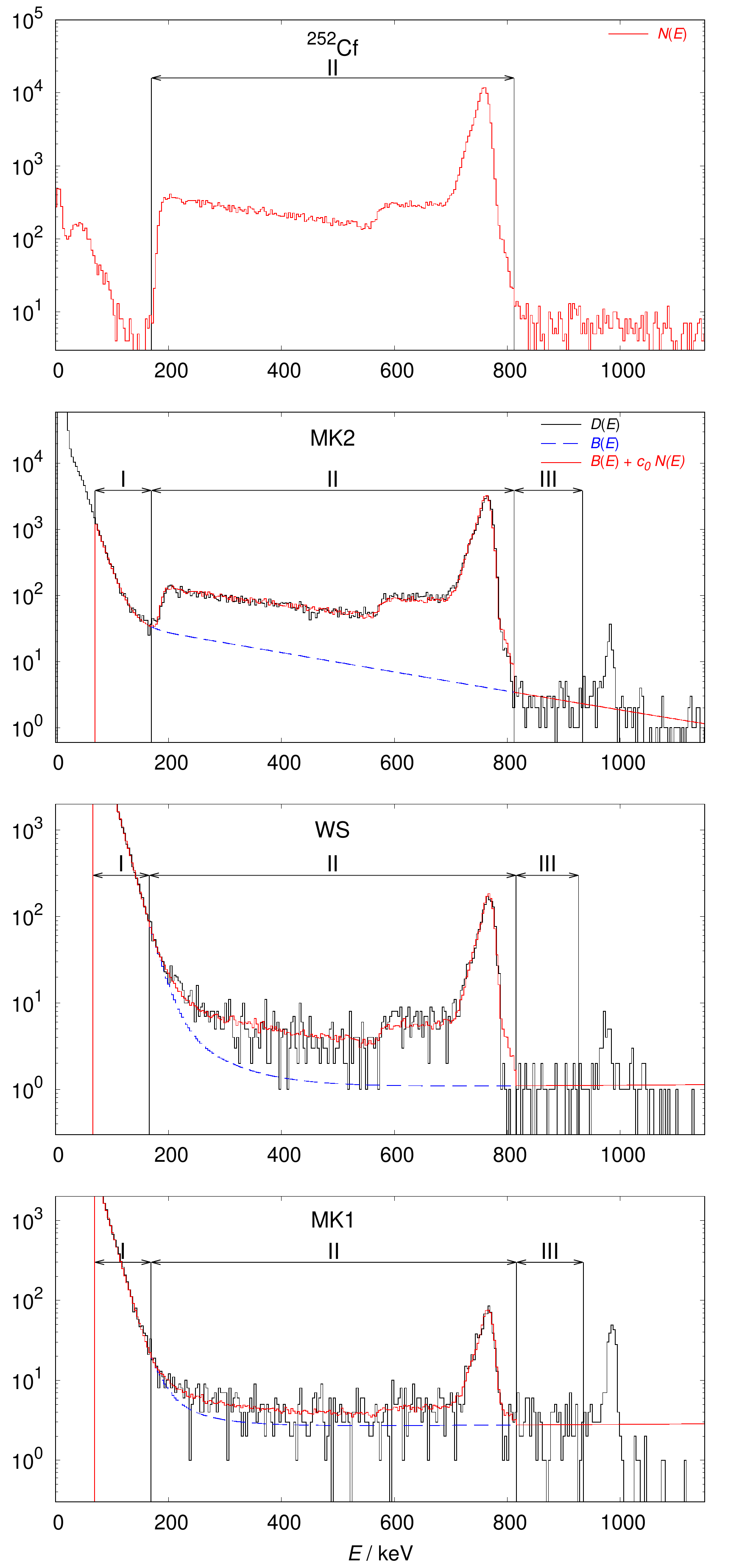}%
\includegraphics[width=0.5\textwidth]{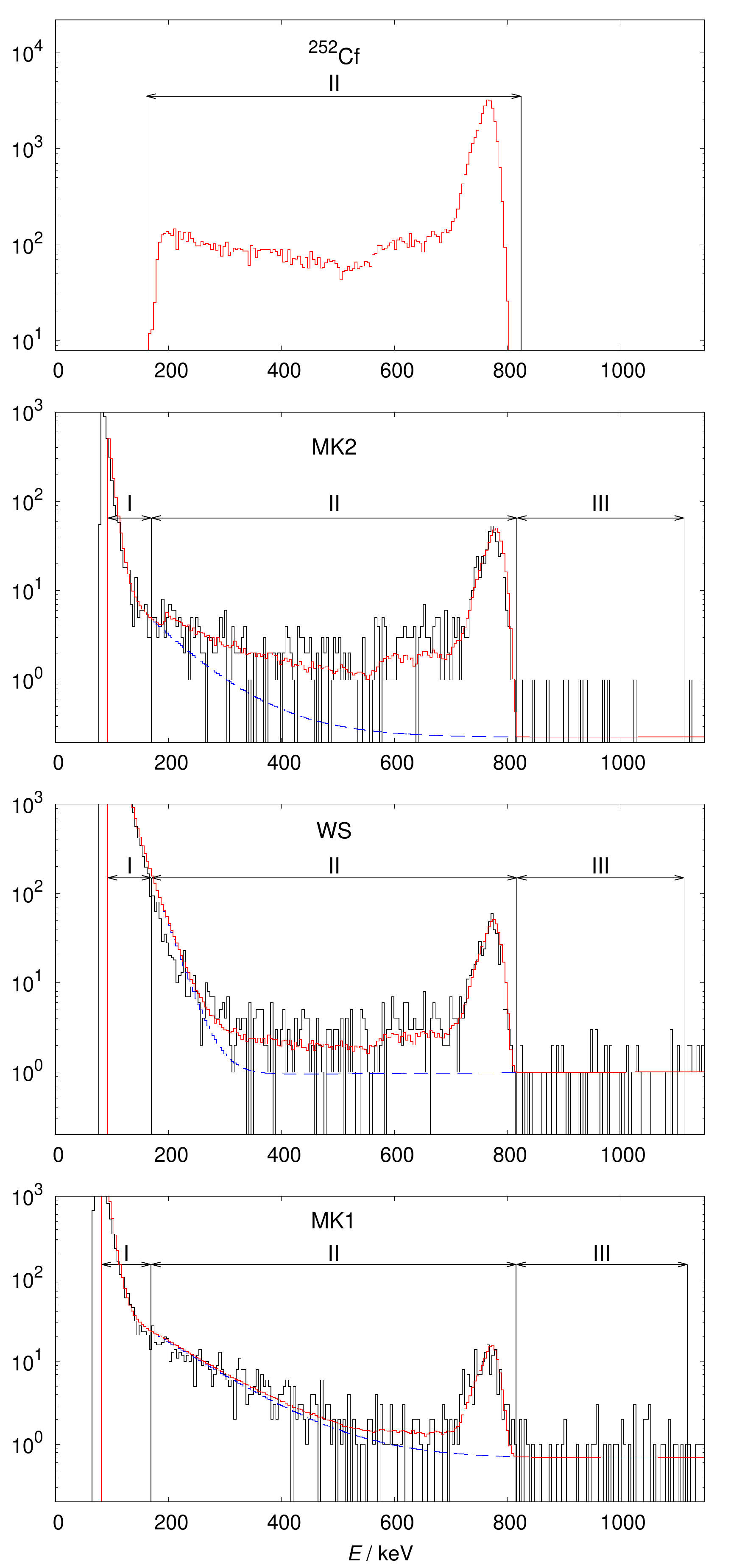}
\caption{\label{fig:Spectrum} Pulse height spectra with detector-moderator assemblies A3 (left, a typical case) and B8 (right, the assembly with the lowest signal-to-noise ratio). Top row, raw data $N(E)$ with $^{252}$Cf source. Rows 2--4 show the neutron flux in MK2 (left: 28.7\,days, right: 3.6 out of 10.6\,days), WS (left: 12.5\,days, right: 14.0\,days), and MK1 (left: 8.6\,days, right: 7.0 out of 25.3\,days), respectively. Raw data $D(E)$ (black histogram), fitted background $B(E)$ (blue dashed curve), and  scaled neutron signal plus fitted background $c_0 N(E)+B(E)$ (red histogram). See text for details.}
\end{figure*}

%%%%%%%%%%%%%%%%%%%%%%%%%%%%%%%%%%%%%%%%%%%%%%%%%%%%%%%%%%%%%%%%%%%%%

%%%%%%%%%%%%%%%%%%%%%%%%%%%%%%%%%%%%%%%%%%%%%%%%%%%%%%%%%%%%%%%%%%%%%
\subsection{Generation of the pulse height spectra}
\label{subsec:SpectrumGeneration}

The shape of the pulse height spectrum in a $^3$He detector is determined by the detection process, which starts with the $^3$He(n,p)$^3$H nuclear reaction. Then, the main peak at 764\,keV is due to the coincident detection of the full energies of both reaction products $p$ and $^3$H in the gas proportional counter. In addition, there are two steps at 191 and 573\,keV due to the so-called wall effect, when either the proton or the triton is not stopped inside the sensitive gas volume, but escapes into the detector walls. 

These three features are clearly visible in the $^{252}$Cf and MK2 spectra (Fig. \ref{fig:Spectrum}, first and second rows). In the WS spectra, only the 573 and 764\,keV features are discernible (Fig. \ref{fig:Spectrum}, third row). For the MK1 case (Fig. \ref{fig:Spectrum}, last row), the 764\,keV peak is always clearly visible but the other features usually not. The lower signal-to-noise ratios for WS and MK1 are due to their lower neutron fluxes.

Given that the $^3$He counters are operated in the proportional regime, the spectra are calibrated linearly in deposited energy $E$ by using the above mentioned features in the $^{252}$Cf runs (Fig. \ref{fig:Spectrum}). 

The pulser used in the underground measurements of campaign A produced a peak slightly above 1000\,keV in the spectrum. For the generation of the pulse height spectra, events with data in all seven channels were assumed to be caused by the pulser and gated out. The remaining pulser feature apparent in Fig.\:\ref{fig:Spectrum} at \mbox{$E$ $\sim$ 1000\,keV} is thus due to nondetection of the pulser signal in another channel and serves to determine the dead time of the DAQ system. 

For the analysis, a region of interest ranging typically from 180 to 820\,keV is adopted (region II in Fig. \ref{fig:Spectrum}), encompassing all three neutron-induced features. It is noted that the detectors are mainly sensitive to thermal neutrons ($\sigma$ = 5333\,barn \cite{Mughabghab06-Book}). Due to the $1/v$ law, the cross section is 1500 times lower for $E_n$ = 56\,keV neutrons, which would register just outside the 820\,keV upper bound of region II. Any significant neutron contribution to the pulse height spectrum outside of region II would thus require a very unlikely neutron energy spectral shape. 
Instead, the remaining continuum at high energy is assumed to be due to intrinsic detector background \cite{Amsbaugh07-NIMA}.

%%%%%%%%%%%%%%%%%%%%%%%%%%%%%%%%%%%%%%%%%%%%%%%%%%%%%%%%%%%%%%%%%%%%%
\subsection{Determination of the neutron count rates}
\label{subsec:ExpCountingrates}

In order to determine the neutron count rates, the observed energy-calibrated pulse height spectrum $D(E)$ in regions I--III (Fig, \ref{fig:Spectrum}) is modeled as the sum of the following components  \cite{Reginatto13-AIPCP}: first, a neutron response $N(E)$ determined in the run with the $^{252}$Cf source with negligible background in region II, scaled to match the measured spectrum, and second, a background term $B(E)$ that is given by 
\begin{equation}
D(E) = c_0 N(E) + B(E) \label{eq:Reginatto}.
\end{equation}

At low pulse heights (region I in Fig. \ref{fig:Spectrum}), the spectrum contains effects of electronic noise which grow exponentially toward the lowest energies and a more slowly background due to residual $\gamma$ rays. These two components are described by the sum of two exponential functions:
\begin{equation}
c_1 \, \exp(-c_2 E) + c_3 \, \exp(-c_4 E).
\end{equation}
At high pulse heights (region III in Fig. \ref{fig:Spectrum}), the intrinsic $\alpha$-activity originating in the housing of the $^3$He gas proportional counter dominates, which is parametrized as
\begin{equation}
c_5 \, (1 + c_6 E)
\end{equation}
with $c_6 \ll$ 1 so that the total background is given by  \cite{Reginatto13-AIPCP}
\begin{equation}
B(E) = c_1 \, \exp(-c_2 E) + c_3 \, \exp(-c_4 E) + c_5 \, (1 + c_6 E).
\end{equation}

For each pulse height spectrum, the parameters $c_i$ ($i \in\{\rm 0,1,...,6\}$) were then fit with the WinBUGS\:1.4 \cite{Winbugs03-Software} Bayesian Markov Chain Monte Carlo algorithm, using the following spectra $D_{j}(E)$ ($j \in\{\rm I, II, III\}$), depending on the region:
\begin{align*}
D_\text{I}(E)   &=                  & c_1 e^{-c_2 E} + &c_3 e^{-c_4 E} + c_5 (1+c_6 E) \\
D_\text{II}(E)  &= c_0 N(E)\, + &c_1 e^{-c_2 E} + &c_3 e^{-c_4 E} + c_5 (1+c_6 E) \\
D_\text{III}(E) &=                  &                      &c_3 e^{-c_4 E} + c_5 \, (1+c_6 E).
\end{align*}

In this way, the rapidly varying part $c_1 e^{-c_2 E}$ of the electronic noise is fitted mainly in region I, and the slowly varying part $c_3 e^{-c_4 E}$ mainly in regions I and III. The $\alpha$-induced background is fitted mainly in region III. The sought after neutron count rate is fitted in region II and encoded in parameter $c_0$. For practical reasons, the fit was performed on the discrete energy bins instead of the energy. 

The resulting fitted background $B(E)$ and modeled total response $B(E) + c_0 N(E)$ are also shown in Fig. \ref{fig:Spectrum}. For detector A3 (Fig. \ref{fig:Spectrum}, left column), the signal-to-noise ratio in  region II is 16.0, 2.8, and 1.3, respectively, for MK2, WS, and MK1. For the worst case, i.e., detector B8, the signal-to-noise ratio is 5.4, 0.6, and 0.3, respectively, in the same three sites.

\subsection{Initial interpretation of the count rate data}
\label{sec:Initial}

%%%%%%%%%%%%%%%%%%%%%%%%%%%%%%%%%%%%%%%%%%%%%%%%%%%%%%%%%%%%%%%%%%%%%
\begin{figure}
\includegraphics[width=\columnwidth,trim=3mm 0 3mm 0]{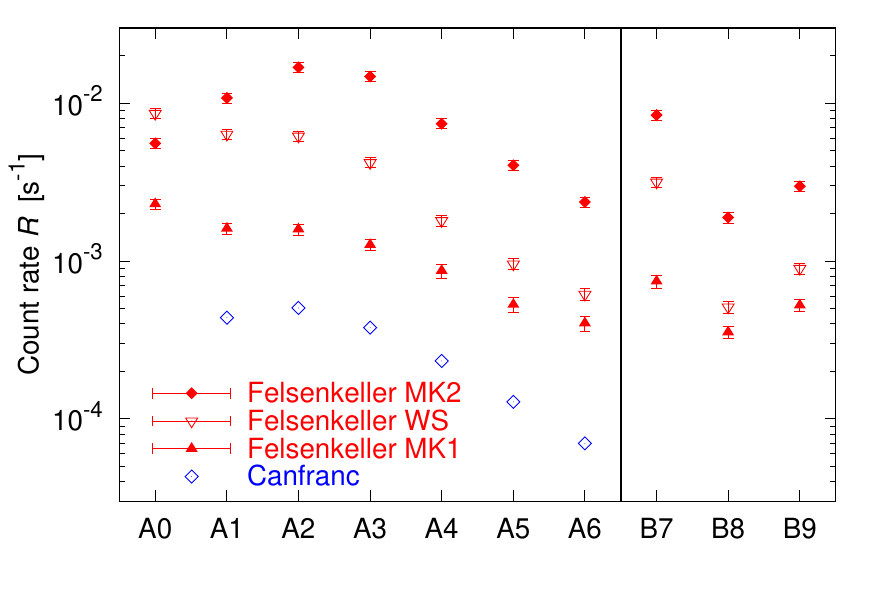}
\caption{\label{fig:Countingrates} Neutron count rates obtained in the present work for Felsenkeller MK2 (red diamonds), WS (red downward triangles), and MK1 (red upwards triangles). For detectors A1--A6, also a comparison with previous rates obtained in Canfranc (blue diamonds, \cite{Jordan13-APP,Jordan13-APP_Corr}) is shown.}
\end{figure}
%%%%%%%%%%%%%%%%%%%%%%%%%%%%%%%%%%%%%%%%%%%%%%%%%%%%%%%%%%%%%%%%%%%%%

The resulting neutron count rates $R_k(x)$ with $x$ the site and $k$ the number of the assembly are shown in Fig. \ref{fig:Countingrates}. The relative uncertainty, as given by the count statistics and the uncertainty on the fitted background, is 2\%--6\% for MK2, 1\%--5\% for WS, and 3\%--9\% (13\% for the case of B8) for MK1.

The count rate data already permit some first observations, prior to further analysis. First, assemblies A1--A6 follow the same general pattern in all three sites studied, and also in the previous Canfranc measurement \cite{Jordan13-APP,Jordan13-APP_Corr}
\begin{equation}
R_k({\rm MK2}) > R_k({\rm WS}) > R_k({\rm MK1}) > R_k({\rm Canfranc}) \nonumber
\end{equation}
for $k \in \{1,...,6\}$. 
Second, the differences between sites MK2 and MK1 are similar to the differences between the lower of the two, MK1, and Canfranc. Third, the thermal neutrons in the unmoderated detector A0 break this pattern and show a higher thermal flux in the unshielded WS than in the heavily shielded MK2.

For MK2, WS, and MK1, the pattern observed for A1--A6 is again evident for B7--B9. The pairs of similar assemblies A2--B7 and A5--B8 show again a similar pattern when MK1 and MK2 are compared. The general patterns of the neutron count rates are therefore consistent across campaigns A and B.

%%%%%%%%%%%%%%%%%%%%%%%%%%%%%%%%%%%%%%%%%%%%%%%%%%%%%%%%%%%%%%%%%%%%%
\subsection{Comparison of $^{252}$Cf source data and FLUKA simulation}
\label{subsec:Cf}

The FLUKA predictions from the FLUKA simulation (Sec. \ref{sec:Sensitivities}) were compared with the data from the $^{252}$Cf source runs. Since the FLUKA code has been extensively validated for its description of neutron interactions \cite{Agosteo12-NIMA}, e.g., the $^{252}$Cf runs serve to verify the correct implementation of detector and moderator geometry in the present simulation.

Here, the $^{252}$Cf source was modeled centrally on top of the detector-moderator assembly in FLUKA. The Mannhart $^{252}$Cf spectrum \cite{IAEA-Tecdoc-410} with a peak at 0.75\,MeV was adopted to describe the neutron emission from the $^{252}$Cf source. Other radiations emanating from the source were neglected, because the $^3$He counter is not very sensitive to $\gamma$ rays. Any detected $\gamma$ rays would form a background that is exponentially decreasing with energy and are subtracted from the experimental signal (see above, Sec. \ref{subsec:ExpCountingrates}).

Assemblies A3 and B7 show the highest overall sensitivities to the $^{252}$Cf neutrons (Fig.\:\ref{fig:Sensitivity}). For these two cases, the simulated count rates are (3$\pm$2)\% and (5$\pm$15)\% lower, respectively, than the relevant experimental rate. The uncertainty is in both cases dominated by the calibration of the $^{252}$Cf source used:  2\% for the source used for A3 (Sec.\:\ref{subsec:CampaignA}) and 15\% for the source used for B7 (Sec.\:\ref{subsec:CampaignB}).

For assemblies A2, A4--A6, and B8--B9, the detector-to-detector ratios of sensitivities are well reproduced, but the simulated rate is up to 10\% below the measured rate. In those cases, due to the mismatch between the $^{252}$Cf spectrum and the spectral sensitivity the observed count rate is influenced by features such as laboratory walls, detector stands, or other detectors, which are imperfectly modeled in the simulation. For assemblies A0--A1, thermal or epithermal neutrons dominate over the emitted 0.75\,MeV neutrons; thus, the $^{252}$Cf data cannot be used.

As a result of these tests, 5\% is adopted as systematic uncertainty for the sensitivity calculated by FLUKA.

%%%%%%%%%%%%%%%%%%%%%%%%%%%%%%%%%%%%%%%%%%%%%%%%%%%%%%%%%%%%%%%%%%%%%
\section{Determination of the experimental neutron flux in Felsenkeller}
\label{sec:Fit}

%%%%%%%%%%%%%%%%%%%%%%%%%%%%%%%%%%%%%%%%%%%%%%%%%%%%%%%%%%%%%%%%%%%%%
\begin{figure}
\includegraphics[width=\columnwidth]{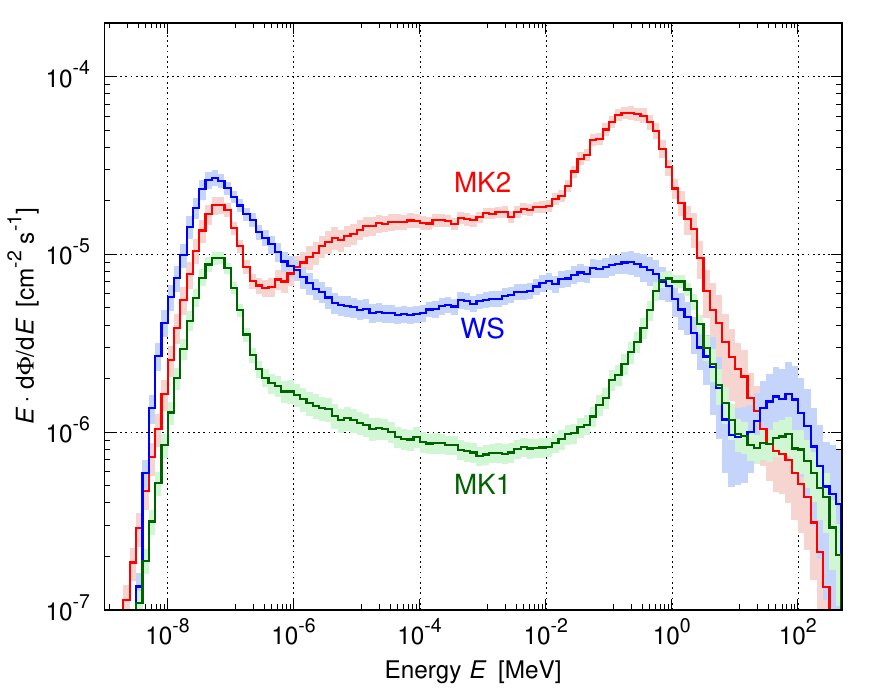}
\caption{\label{fig:All_spectra} Unfolded spectra using the MAXED code, for MK2\:(red), WS\:(blue), and MK1\:(green). The $1\sigma$ error bars are shown as shaded areas; see text for details.
}
\end{figure}
%%%%%%%%%%%%%%%%%%%%%%%%%%%%%%%%%%%%%%%%%%%%%%%%%%%%%%%%%%%%%%%%%%%%%

The measured count rates (Sec. \ref{subsec:ExpCountingrates}) and calculated sensitivities (Sec. \ref{sec:Sensitivities}) were used to determine the neutron energy spectrum and flux. The results are shown in this section.

\subsection{General approach for the fit}
\label{subsec:FitApproach}

As a first step, for each detector $i$, the integral in Eq.~(\ref{eq:SpectralSensitivity}) is approximated by a sum, using the calculated spectral sensitivities (Fig. \ref{fig:Sensitivity}, Sec. \ref{sec:Sensitivities}) in the 396 energy bins given by FLUKA,
\begin{equation} \label{eq:SpectralSensitivitySum}
R_i^\text{exp} = \sum\limits_{j=1}^{396} S_i(E_j) \Phi(E_j)  .
\end{equation}
Here, $R_i^\text{exp}$ with $i\in\{0,...,9\}$ is the experimental count rate in assembly $i$, $S_i(E_j)$ the calculated sensitivity for assembly $i$ and central energy $E_j$ of energy bin $j$ for \mbox{$j\in\{1,...,396\}$}, and $\Phi(E_j)$ the sought after neutron flux in the same energy bin. This can also be expressed as a matrix equation

\begin{equation}
\begin{pmatrix}
R_0^\text{exp} \\
\cdots{} \\
R_9^\text{exp} \\
\end{pmatrix}
= 
\begin{pmatrix}
S_0(E_1) & \cdots & S_0(E_{396}) \\
\cdots & \cdots & \cdots \\
S_9(E_1) & \cdots & S_9(E_{396}) \\
\end{pmatrix}
\times
\begin{pmatrix}
\Phi(E_1) \\
\cdots{} \\
\Phi(E_{396}) \\
\end{pmatrix}.
\end{equation}

For the fit to determine $\Phi(E_j)$, this linear inverse problem must be solved. To this end, subsequently two different codes called MAXED \cite{Reginatto02-NIMA} and GRAVEL \cite{Matzke94-Gravel}, respectively, were used. Both codes are included in the Nuclear Energy Agency's UMG 3.3 package \cite{NEA04-UMG}. 
{\it Nota bene} these codes do not give exact mathematical solutions to problem (\ref{eq:SpectralSensitivitySum}), because there are only ten measured count rates $R_i^\text{exp}$, one for each detector assembly.

Therefore, the solution space is limited by using physically motivated {\it a priori} spectra as starting points of the fit. Here, for each of the three sites studied, the respective predicted spectrum $\Phi^\text{prior}(E)$ (Sec. \ref{sec:PredictedFlux}) was taken as a starting point. Both codes then derive a new spectrum based on the starting point and on the measured data. 

The first code used here, MAXED, performs the fit by maximizing the entropy function \cite{Reginatto02-NIMA}:
\begin{align}
	S = - \sum\limits_{j=1}^{396} \left( \Phi(E_j)\, \ln \frac{\Phi(E_j)}{\Phi^\text{prior}(E_j)} + \Phi^\text{prior}(E_j) - \Phi(E_j) \right).
\end{align}
In order to remain close to a physically reasonable scenario, the solution is constrained by a limit called $\Omega$ on the $\chi^2$ parameter, 
\begin{align}
	\Omega \stackrel{!}{\geq}  \chi^2 = \sum\limits_{i=0}^{9} \left( \frac{R_i^\text{calc}-R_i^\text{exp}}{\Delta R_i^\text{exp}} \right)^2 \label{eq:ChiSquare}
\end{align}
where $R_i^\text{calc}-R_i^\text{exp}$ is the difference between the calculated and observed count rates for detector $i$ and $\Delta R_i^\text{exp}$ is the experimental uncertainty of $R_i^\text{exp}$. 
$\Omega$ is usually set equal to the number of detectors. For the present purposes, solutions with $\Omega=10$--16 were used. 
In the fit process, it is assumed that the errors of $R_i^\text{exp}$ are normally distributed with zero mean and variance $(\Delta R_i^\text{exp})^2$.

The second code, GRAVEL, uses a slightly modified version of the SAND-II code \cite{McElroy67-Book} and works iteratively. Based on the calculated sensitivities $S_i(E_j)$ and the \mbox{\it a priori} spectrum $\Phi^\text{prior}(E)$, the expected neutron rates  $R_i^\text{calc}$ are calculated and compared to the measured rates  $R_i^\text{exp}$ to provide a correction factor $f_i$ for each detector $i$. 

For each of the 396 energy bins $j$, the correction factor $f_i$ is then weighted by the detector sensitivity $S_i(E_j)$ for this detector and energy bin and is applied. The resulting spectrum is then used as the starting spectrum for the next iteration. The iteration process stops once the requested value $\Omega$ was reached.

%%%%%%%%%%%%%%%%%%%%%%%%%%%%%%%%%%%%%%%%%%%%%%%%%%%%%%%%%%%%%%%%%%%%%
\subsection{Extracted neutron fluxes and their uncertainties}

%%%%%%%%%%%%%%%%%%%%%%%%%%%%%%%%%%%%%%%%%%%%%%%%%%%%%%%%%%%%%%%%%%%%%
\begin{figure}[tb]
\includegraphics[width=\columnwidth,trim=5mm 0 5mm 4mm]{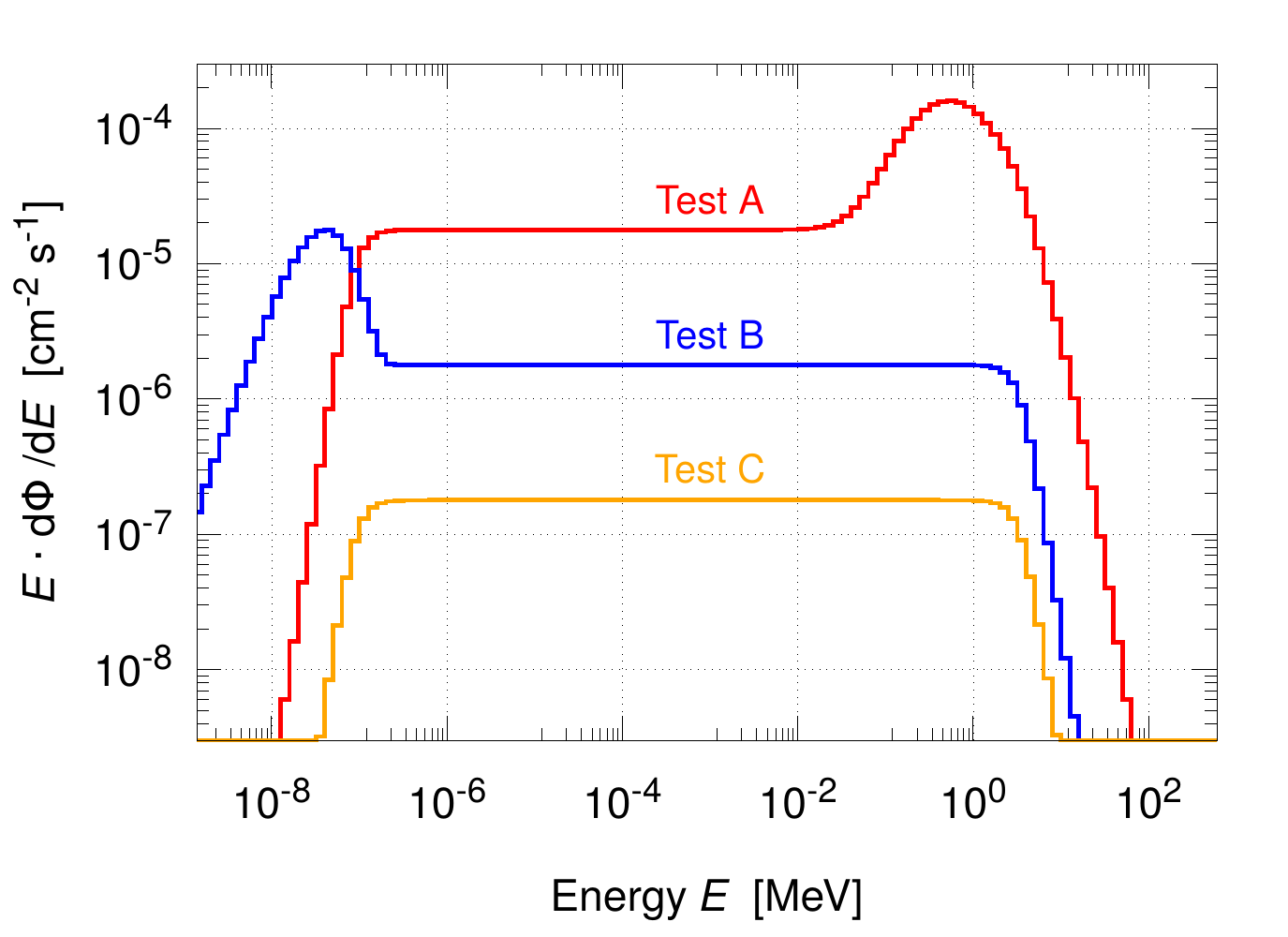}
\caption{\label{fig:TestSpectra} The three extreme-case test spectra called A, B, and C used for the error analysis. See text for details.}
\end{figure}
%%%%%%%%%%%%%%%%%%%%%%%%%%%%%%%%%%%%%%%%%%%%%%%%%%%%%%%%%%%%%%%%%%%%%

The measured neutron fluxes as extracted with MAXED are shown in Fig. \ref{fig:All_spectra}. For both MAXED and GRAVEL, the integral flux is listed in Table \ref{Table:Flux}. 

The error bands in the energy spectra (shaded regions in Fig. \ref{fig:All_spectra}) were obtained by using the IQU software that is contained within the UMG~3.3 package. IQU considers variations of the measured data, quantified by their quoted uncertainty, and performs a sensitivity analysis and uncertainty propagation \cite{NEA04-UMG}.

For all three locations, the relative errors determined by IQU are 10\%--15\% for energies below 10\,MeV, and 3--4 times higher above 10~MeV. This increase in uncertainty is due to the declining overall neutron sensitivity (Fig. \ref{fig:Sensitivity}) and a resulting decline in detected neutron events, and also due to the fact that the sensitivity of the lead-lined moderator keeps increasing toward higher energies and, thus, cannot be normalized. 

Since the IQU error determination is only available for MAXED and depends on the initial unfolding spectrum, an even more conservative approach was used to calculate the uncertainties of the integrated flux values $\Phi_\text{exp}$ shown in Table \ref{Table:Flux}. In order to exclude a possible bias due to the characteristics of the FLUKA-derived predicted spectra (Fig. \ref{fig:SimulatedSpectra}), three hand-designed extreme test spectra were used: an ($\alpha,n$)-dominated (Test A), a thermal-dominated (Test B), and a flat spectrum \mbox{(Test C)}, all three shown in Fig. \ref{fig:TestSpectra}. These three spectra should not be viewed as real physics cases but as extreme bounds encompassing all plausible physical solutions.

Using the spectra Test A, B, and C, the unfolding was repeated with both MAXED and GRAVEL, resulting in integrated test fluxes $\Phi_i$ ($i \in \{\text{A,B,C}\}$). The error $\Delta\Phi_\text{exp}$ was then taken to be $\Delta\Phi_\text{exp} = \sqrt{\sum_{i}(\Phi_i-\Phi_\text{exp})^2/3}$. This procedure results in 7\%--9\% uncertainty for the integrated flux. If one were to use only the IQU errors instead, a much lower uncertainty of typically 1\% would be found, close to the combined statistical uncertainty of the count rates. Thus, the above described and adopted approach with the test spectra is conservative.

When adding the 5\% uncertainty adopted from the $^{252}$Cf test (Sec. \ref{subsec:Cf}) and 4\% uncertainty from angular effects (Sec. \ref{subsec:Anisotropy}), the final flux uncertainty is 9\%--11\% (Table \ref{Table:Flux}).

%%%%%%%%%%%%%%%%%%%%%%%%%%%%%%%%%%%%%%%%%%%%%%%%%%%%%%%%%%%%%%%%%%%%%
\begin{table*}
\centering
\begin{tabular}{|l|llll|lll|}
\hline
~~Location 	 & \multicolumn{4}{c|}{Predicted flux [10$^{-4}$\,cm$^{-2}$\,s$^{-1}$]}	& \multicolumn{3}{c|}{~~Measured flux [10$^{-4}$\,cm$^{-2}$\,s$^{-1}$]~~}  \\
~~inside		 	& \multicolumn{3}{c}{All energies} & \multicolumn{1}{c|}{~~10--300\,MeV}  & \multicolumn{2}{c}{All energies} & \multicolumn{1}{c|}{10--300\,MeV~~} \\ \cline{2-5}\cline{6-8}
~~Felsenkeller~~		 	& ~~($\mu,n$)		& \multicolumn{1}{c}{($\alpha,n$)~~~}	& ~~Total	& \multicolumn{1}{c|}{total} & \multicolumn{1}{l}{~~MAXED~~} &  ~~GRAVEL~~ &  ~~MAXED~~		\\
\hline \hline
~~MK2      	& ~~5.8\,(4)	& ~~0.032\,(8)         & ~~5.8\,(4)	  & \multicolumn{1}{c|}{0.040\,(3)}	& ~~~4.6\,(4)   & ~~~~4.6\,(4)     & \multicolumn{1}{c|}{0.04\,(2)}    \\
~~WS    	& ~~1.65\,(13)	& ~~0.20\,(5)          & ~~1.85\,(13) & \multicolumn{1}{c|}{0.029\,(3)}	& ~~~1.96\,(15) & ~~~~$2.00\,(16)$ & \multicolumn{1}{c|}{0.04\,(2)}    \\
~~MK1      	& ~~0.50\,(4)   & ~~$Out: 0.013\,(3)$  & ~~0.51\,(4)  & \multicolumn{1}{c|}{0.021\,(2)}	& ~~~0.61\,(5)  & ~~~~$0.63\,(6)$  & \multicolumn{1}{c|}{0.03\,(1)}    \\
            &               & ~~$In: 0.00007\,(2)$ & & & & &\\ \hline
\end{tabular}
\caption{Neutron flux in the three sites MK2, WS, and MK1 inside Felsenkeller tunnel IV. Uncertainties are shown in parentheses. The predicted fluxes (Sec.\:\ref{sec:PredictedFlux}) are compared with the deconvoluted measured fluxes. The data are shown separately for all energies (10$^{-9}$--300\,MeV) and only for the high-energy segment. The ``out'' location in MK1 corresponds to the location of detectors A0--A3 (Fig. \ref{fig:Locations}).}
\label{Table:Flux}
\end{table*}
%%%%%%%%%%%%%%%%%%%%%%%%%%%%%%%%%%%%%%%%%%%%%%%%%%%%%%%%%%%%%%%%%%%%%

%%%%%%%%%%%%%%%%%%%%%%%%%%%%%%%%%%%%%%%%%%%%%%%%%%%%%%%%%%%%%%%%%%%%%
\begin{table}
\centering
\begin{tabular}{|l|l|D{.}{.}{-1}|}
\hline
Location & Reference & \multicolumn{1}{c|}{Flux} \\ %\cline{1-6} 
		&		& \multicolumn{1}{c|}{[10$^{-4}$\,cm$^{-2}$\,s$^{-1}$]} \\ \hline
Ground level & \cite{Wiegel02-NIMA_Surface} & \multicolumn{1}{D{.}{}{-1}|}{121.\,(6)} \\ %\hline
YangYang, 2000\,m.w.e. & \cite{Park13-Apradiso} & 0.67\,(2) \\ %\hline
Canfranc, 2400\,m.w.e. (revised) & \cite{Jordan13-APP,Jordan13-APP_Corr} &  0.138\,(14) \\ %\hline
Gran Sasso, 3800\,m.w.e. & \cite{Belli89-NCA} & 0.038\,(2) \\ \hline
Felsenkeller MK2, 140\,m.w.e. & Present & 4.6\,(4) \\
Felsenkeller WS, 140\,m.w.e. & Present & 1.96(15) \\
Felsenkeller MK1, 140\,m.w.e. & Present & 0.61(5) \\ \hline
\end{tabular}
\caption{Measured energy-integrated neutron fluxes from the literature \cite{Wiegel02-NIMA_Surface,Park13-Apradiso,Jordan13-APP,Jordan13-APP_Corr,Belli89-NCA} and from the present work.}
\label{Table:FluxAllLabs}
\end{table}

%%%%%%%%%%%%%%%%%%%%%%%%%%%%%%%%%%%%%%%%%%%%%%%%%%%%%%%%%%%%%%%%%%%%%
\section{Discussion}
\label{sec:Discussion}

\subsection{Experimental neutron flux at Felsenkeller}

The count rates of the individual detectors \mbox{(Sec.\:\ref{sec:Initial},} Fig.\:\ref{fig:Countingrates}) show a general pattern that is similar for all three sites studied and also for the previous measurement in Canfranc, Spain \cite{Jordan13-APP,Jordan13-APP_Corr}. There is a significant overall difference, by roughly a factor of 7, between the count rates observed in sites MK1 and MK2. The relative differences are largest for the assemblies moderated by 7--18\,cm polyethylene, which are mainly sensitive in the neutron energy range from 10$^{-6}$ to 1\,MeV, and smallest for the unmoderated detector. 

This overall trend between the three sites MK2, WS, and MK1 had already been observed previously in a study with two moderated $^3$He counters, both of them with an energetic response similar to the present assembly A3 \cite{Niese07-JRNC}. 

The neutron fluxes and energy spectra resulting from the deconvolution algorithm (Fig. \ref{fig:SimExp}) are again quite different between the three particular sites studied in Felsenkeller, again with the lowest flux for MK1, the highest for MK2, and WS in between. 

%%%%%%%%%%%%%%%%%%%%%%%%%%%%%%%%%%%%%%%%%%%%%%%%%%%%%%%%%%%%%%%%%%%%%
\begin{figure}[tb]
\includegraphics[width=\columnwidth]{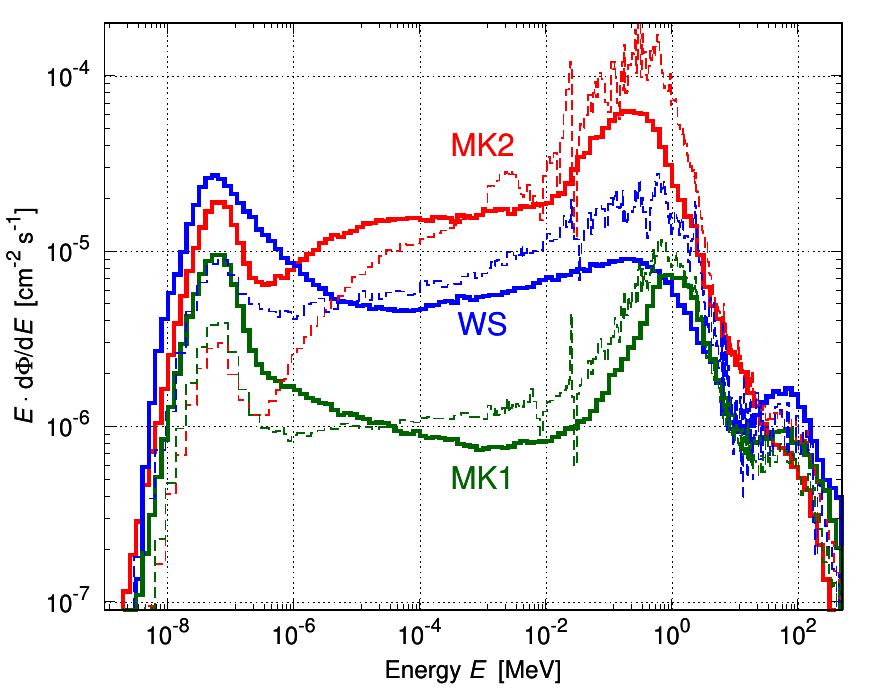}
\caption{%
\label{fig:SimExp} 
Comparison of the experimental (full lines) and predicted (dashed lines) neutron spectra in the three Felsenkeller sites MK2 (red), WS (blue), and MK1 (green). See text for details.
}
\end{figure}
%%%%%%%%%%%%%%%%%%%%%%%%%%%%%%%%%%%%%%%%%%%%%%%%%%%%%%%%%%%%%%%%%%%%%

\subsection{Comparison of data and simulation}
\label{subsec:Comparison_data_simulation}

The energy-integrated neutron fluxes predicted by FLUKA follow the trend of the data (Table \ref{Table:Flux}), but they show relative differences of +(26$\pm$10)\%, --(6$\pm$10)\%, and --(16$\pm$10)\%, for MK2, WS, and MK1, respectively.

When comparing predicted and experimental energy spectra (Fig. \ref{fig:SimExp}), it is seen that in all cases, the simulation overpredicts the intermediate to fast energy range \mbox{($E$ = 10$^{-5}$--10\,MeV)}. In contrast, the thermal neutron flux ($E$ $\sim$ 2.5$\times$10$^{-8}$\,MeV) is always underpredicted. These two effects may be due to existing low-density materials such as plexiglass housings for detectors, wooden tables and shelves, and the liquid nitrogen in the dewars of the HPGe\:detectors that were all neglected in the FLUKA simulation. These materials may moderate higher energy neutrons to lower energies, explaining the disequilibrium between the two above mentioned energy ranges. In addition, thermal neutrons are more likely to be absorbed by structural materials, which may explain some of the overprediction of the observed energy-integrated flux in MK2 and WS.

When considering MK1, where also the energy-integrated flux is somewhat underpredicted, it is noted that for this case of generally very low ($\mu,n$) production, even limited quantities of neglected materials may enhance neutron production. In addition, due to space constraints, the smaller detectors A0--A3 had to be placed in a part of MK1 surrounded only on five out of six sides by serpentinite, with the sixth side showing WS shielding conditions.

For the case of WS, the data show a pronounced thermal peak that is not present in the simulation. This may in principle be caused by moderation in the humid rock walls \cite{Wulandari04-APP}, where the present FLUKA simulation assumed 3\% water content (by mass). Since the true humidity changes with weather conditions, i.e. air temperature, humidity and precipitation of the preceding days, it is not easy to model it without daily {\it in situ} measurements. Another possibility, at least in principle, are thermal neutrons leaking in through the doors of the laboratory.

In order to assess the effects of humidity, FLUKA simulations with varying water content were performed for all three sites ranging from 3\% (adopted value) up to 12\% (extreme case). For 12\% water content the integrated predicted fluxes change by $-13$\% (WS), not at all (inside MK1), and $-7$\% (MK2). 

In the energy region from 10 to 300\,MeV, due to the limited statistics and the low observed flux, the flux data are only 2--3$\,\sigma$ above zero (Table \ref{Table:Flux}). The prediction is consistent with these limited-precision data. 

When considering the matching between simulation and underground neutron flux data, it is important to note that it was recently reported that GEANT4 simulations underpredicted the flux of ($\mu,n$) neutrons \cite{Du18-APP} in a laboratory with just 13\,m.w.e. rock overburden. More recent work by the same group also included FLUKA simulations and showed a good match of simulation and data \cite{Kneissl19-APP}. Earlier work in deep-underground settings reported a good match between simulation and data both for GEANT4\:\cite{Zhang14-PRD} and FLUKA\:\cite{Empl14-JCAP}. The present data suggest a reasonable description of the ($\mu,n$) flux by FLUKA at 140\,m.w.e.

%%%%%%%%%%%%%%%%%%%%%%%%%%%%%%%%%%%%%%%%%%%%%%%%%%%%%%%%%%%%%%%%%%%%%
\begin{figure}[tb]
\includegraphics[width=\columnwidth,trim=3mm 0 6mm 1.5mm]{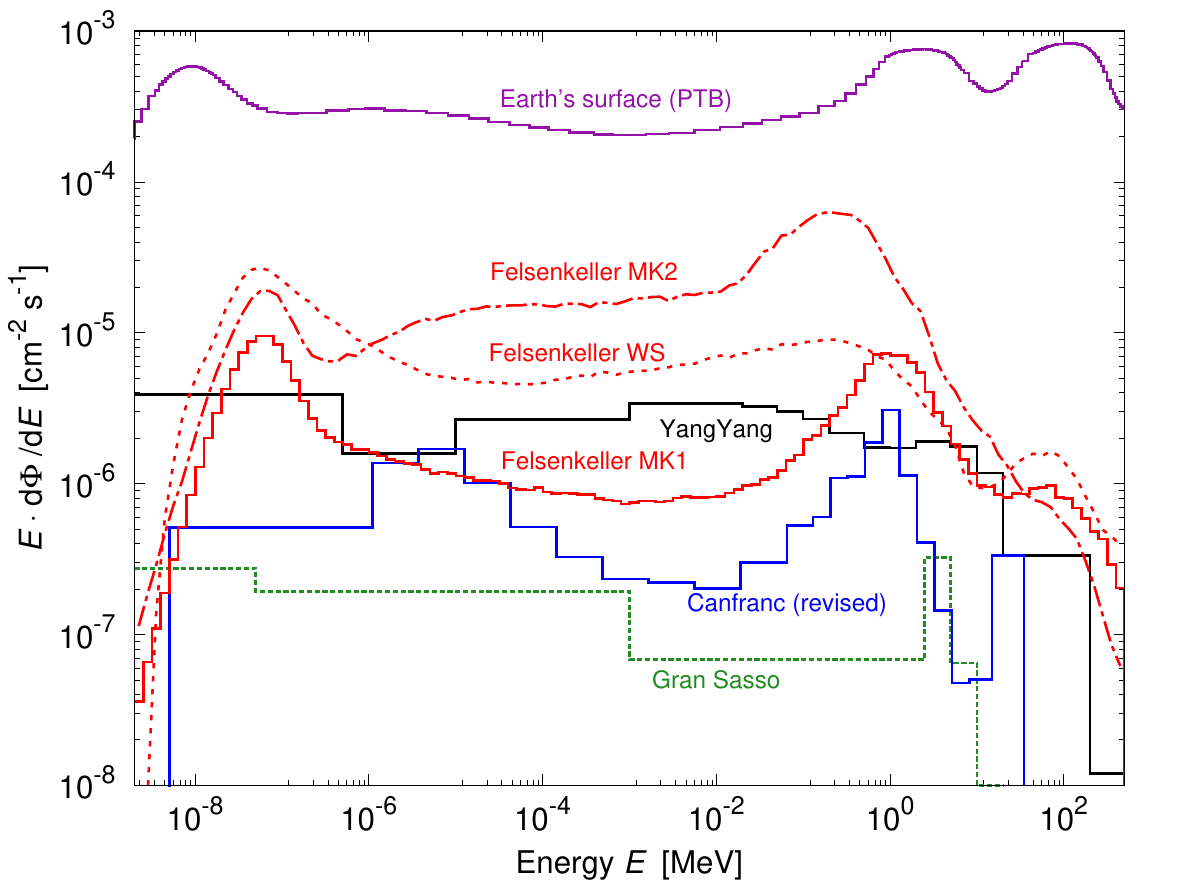}
\caption{%
\label{fig:ExperimentalSpectrum} 
Unfolded neutron flux from the present work, for Felsenkeller tunnel\:IV MK1 (red solid line), MK2 (red dot-dashed line), WS (red dashed line), and the revised Canfranc flux (blue solid line, \cite{Jordan13-APP,Jordan13-APP_Corr}). For comparison, the plot includes previous spectra from the Earth's surface (purple line, \cite{Wiegel02-NIMA_Surface}), YangYang/South Korea (black line \cite{Park13-Apradiso} and Gran Sasso (green dotted line \cite{Belli89-NCA}).
}
\end{figure}
%%%%%%%%%%%%%%%%%%%%%%%%%%%%%%%%%%%%%%%%%%%%%%%%%%%%%%%%%%%%%%%%%%%%%

\subsection{Neutron flux at various underground sites}

The energy-dependent neutron flux from the present work is compared to previous measured neutron spectra in other sites, using a logarithmic presentation  (Fig. \ref{fig:ExperimentalSpectrum}). For the same sites, the integrated fluxes are shown in Table \ref{Table:FluxAllLabs}. 

For the Earth's surface, data by the PTB NEMUS group are used \cite{Wiegel02-NIMA_Surface}. There, the 100\,MeV peak is more than 3 orders of magnitude stronger than at Felsenkeller. Since the Felsenkeller muon flux is only 40 times lower than at the Earth's surface \cite{Olah16-NPA6,Ludwig17-Master,Ludwig19-APP}, the remainder of the difference is presumably due to neutrons produced in the atmosphere, which are completely absorbed at Felsenkeller depth. 

Now, the present data are briefly compared to previous neutron spectrum measurements in deep-underground laboratories. 
The first example, the YangYang laboratory in South Korea, is located below 700\,m of rock (2000\,m.w.e.). There, the neutron flux was studied using a Bonner sphere spectrometer modeled on the PTB NEMUS system \cite{Wiegel02-NIMA}, including several modified spheres with neutron multipliers  \cite{Park13-Apradiso}. The YangYang spectrum is much flatter than the present one, and the downscattered neutron peak at 0.3\,MeV is not evident at YangYang. It is interesting that the reported total flux at YangYang is similar to the present MK1 result, even though the depth is much greater and the reported $^{238}$U and $^{232}$Th contents are rather low, 6--23\,Bq/kg \cite{Lee06-PLB}. 

The spectrum in the Canfranc underground laboratory, Spain (2400\,m.w.e.), is obtained by multiplying the original data \cite{Jordan13-APP} by a factor of 4 \cite{Jordan13-APP_Corr}. The Canfranc spectrum has a similar structure as MK2, but a lower overall flux. Neither a bare nor a lead-lined $^3$He counter was used at Canfranc. As a consequence, the flux obtained at thermal energies and above 10\,MeV are affected by large uncertainties, as discussed in Ref.\:\cite{Jordan13-APP}.

The Canfranc flux, after revision \cite{Jordan13-APP,Jordan13-APP_Corr}, is somewhat higher than expected for its depth of 2400\,m.w.e. This may in principle be due to a very high $^{238}$U/$^{232}$Th content in the rock or due to a different rock chemical composition leading to a more efficient ($\alpha,n$) process. Still, it seems advisable to remeasure the spectral neutron flux at this site.

For Gran Sasso (3800\,m.w.e.), Belli {\it et al.} \cite{Belli89-NCA} spectrum is shown (Fig. \ref{fig:ExperimentalSpectrum}).  
The Gran Sasso flux, similar to YangYang, shows much less structure than Canfranc or Felsenkeller, in that case possibly due to the limited number of energy bins used. In the future, it would be interesting to obtain better data, in particular near 0.3\,MeV.

%%%%%%%%%%%%%%%%%%%%%%%%%%%%%%%%%%%%%%%%%%%%%%%%%%%%%%%%%%%%%%%%%%%%%
\section{Summary and conclusions}
\label{sec:Conclusion}

Using two sets of altogether nine moderated $^3$He counters and one unmoderated $^3$He counter, the neutron flux and spectrum were investigated in three sites in tunnel IV of the Felsenkeller underground facility, Dresden, Germany. The resulting energy-integrated fluxes were ($0.61 \pm 0.05$), ($1.96 \pm 0.15$), and ($4.6 \pm 0.4) \times 10^{-4}$\,cm$^2$\,s$^{-1}$, for sites MK1, WS, and MK2, respectively. 

The data are matched reasonably well by a detailed FLUKA simulation taking into account the known muon flux and angular distribution and the known specific radioactivity of the rock. 

In view of the crucial importance of a proper understanding of the underground neutron background, it seems advisable to reinvestigate the underground neutron flux and energy spectrum at several other sites, including deep-underground laboratories.

The present data were instrumental in the planning for the new Felsenkeller underground ion accelerator laboratory, located in tunnels\:VIII and IX of the same underground site studied here \cite{Bemmerer18-SNC}. In particular, the shielding for the new laboratory was designed to resemble the lowest neutron flux site found here, MK1. Neutron background data from the new facility will be reported in due course.

%%%%%%%%%%%%%%%%%%%%%%%%%%%%%%%%%%%%%%%%%%%%%%%%%%%%%%%%%%%%%%%%%%%%%
\section*{ACKNOWLEDGMENTS}

The authors are indebted to Alfredo Ferrari (CERN) for helpful discussions regarding the FLUKA code. ---
Financial support by DFG\:\mbox{(BE 4100/4-1)}, the Helmholtz Association (NAVI \mbox{VH-VI-417} and \mbox{ERC-RA-0016}), the Spanish Ministerio de Econom\'ia y Competitividad  (Grants No. FPA2014-52823-C2, No. FPA2017-83946-C2, No. RTI2018-098868-B-I00, and No. SEV-2014-0398 / program Severo Ochoa), and the COST Association (ChETEC CA16117) is gratefully acknowledged.

\end{document}
%%%%%%%%%%%%%%%%%%%%%%%%%%%%%%%%%%%%%%%%%%%%%%%%%%%%%%%%%%%%%%%%%%%%%